\documentclass[12pt]{article}
\usepackage{amsfonts}
\usepackage[dvips]{color}
\usepackage{psfrag}
\usepackage{feynmp}
\usepackage{graphicx}
\usepackage{braket}
\usepackage{bm}
\usepackage{subfigure}

\usepackage[normalem]{ulem}
\usepackage{comment}
\includecomment{pdffig}

\usepackage[height=8.5in,width=6.4in]{geometry} 
\usepackage{xparse}
\usepackage{xcolor}
\usepackage{tikz}
\usepackage{cite}
\usepackage[linktocpage]{hyperref}
\usepackage{amssymb, mathtools}
\usepackage[vcentermath]{youngtab}

\newcommand{\cA}{\mathcal{A}}
\newcommand{\cI}{\mathcal{I}}
\newcommand{\PE}{\operatorname{PE}}
\newcommand{\Tr}{\operatorname{Tr}}
\newcommand{\adj}{\mathrm{adj}}
\newcommand{\so}{\mathfrak{so}}
\newcommand{\slc}{\mathfrak{sl}}
\newcommand{\su}{\mathfrak{su}}
\newcommand{\V}{\mathbb{V}}
\newcommand{\C}{\mathbb{C}}
\newcommand{\Z}{\mathbb{Z}}

\usetikzlibrary{cd}

\allowdisplaybreaks[1]

\Yboxdim{5pt}

\makeatletter
    
    \@addtoreset{equation}{section}
  \makeatother

\def\rem#1{}

\renewcommand{\title}[1]{\vbox{\center\LARGE{#1}}\vspace{5mm}}
\renewcommand{\author}[1]{\vbox{\center\large#1}\vspace{5mm}}

\numberwithin{equation}{section}

\DeclareFontShape{OT1}{cmr}{mx}{n}%
    {<->cmr10}{}

\allowdisplaybreaks

\begin{document}
\bibliographystyle{utphys}

 \begin{titlepage}
 \begin{center}
 \vspace{5mm}
 \hfill {\tt 
 }\\
 \vspace{14mm}

 \title{
 \LARGE  
SCFT/VOA correspondence and R-twisted reductions of
$(A_2,D_{3n-2})$ Argyres--Douglas theories}
 \vspace{7mm}

 Yutaka Yoshida

 \vspace{6mm}

 \vspace{3mm}
 {\small {\it Department of Current Legal Study, Faculty of Law,  Meiji Gakuin University, 1-2-37 
 Shirokanedai, Minato-ku, Tokyo 108-8636, Japan}} \\
 {\small {\it Institute for Mathematical Informatics, Meiji Gakuin University,
 1518 Kamikurata-cho, Totsuka-ku, Yokohama, Kanagawa 244-8539, Japan}}

 \end{center}

 \vspace{7mm}
\abstract{
We study the SCFT/VOA correspondence for the $(A_2,D_{3n-2})$
Argyres--Douglas theories with $n\geq2$. From the matching of the central
charges and the equality of the Schur index with the vacuum supercharacter
to all orders, we propose that the associated VOA is the logarithmic doublet
algebra $\mathcal A(4n-2)$ of Feigin, Feigin, and Tipunin. For $n=2$, this
correspondence was studied in detail by Buican and Nishinaka, and our proposal
extends it to $n\geq3$. Using the Nahm sum expression for the vacuum
supercharacter, we further propose a family of 3d $\mathcal N=2$ abelian
Chern--Simons matter theories that flow to the 3d $\mathcal N=4$ SCFTs
obtained by the R-twisted circle reduction, generalizing the $n=2$
construction. We determine
the monopole superpotential and show that the Coulomb branch is
$\mathbb C^2/\mathbb Z_2$ for every $n$. The superconformal index provides
quantitative evidence for the proposed infrared fixed points, including the
expected enhancement to $\mathcal N=4$ supersymmetry and the Higgs and
Coulomb branch structure.
}
\vfill

 \end{titlepage}

\tableofcontents

\section{Introduction}
\label{sec:introduction}

Protected sectors provide one of the few settings in which strongly coupled
four-dimensional (4d) quantum field theories can be studied exactly.  For a
4d $\mathcal N=2$ superconformal field theory, the cohomology of
Schur operators carries the structure of a two-dimensional (2d) vertex operator
algebra (VOA), and the Schur index agrees with its  vacuum
supercharacter \cite{Beem:2013sza}.  This correspondence converts protected
operator data of a generally non-Lagrangian theory into 2d
representation theoretic data.  Argyres--Douglas theories were originally
discovered as strongly coupled fixed points with mutually nonlocal massless
states and were subsequently generalized through irregular singularity and
related constructions \cite{Argyres:1995jj, Argyres:1995xn, Eguchi:1996vu, Cecotti:2010fi, Xie:2012hs,  Wang:2015mra}.
They are particularly well suited to the SCFT/VOA correspondence: their
protected quantities often admit explicit expressions that can be compared
with characters of non-unitary and logarithmic VOAs
\cite{Buican:2016arp}.

The first question addressed in this paper is the Schur sector VOA of the
infinite family
\begin{equation}
 (A_2,D_{3n-2}),
 \qquad n\geq2.
 \label{eq:intro-AD-family}
\end{equation}
The first member, $(A_2,D_4)$, is known to be associated with the logarithmic
doublet algebra $\cA(6)$ \cite{Buican:2016arp}. 
We propose that the VOA associated with the $(A_2,D_{3n-2})$ Argyres--Douglas theory is
\begin{equation}
 \V\bigl((A_2,D_{3n-2})\bigr)
 \simeq
 \cA(4n-2).
 \label{eq:intro-main-correspondence}
\end{equation}
For $n>2$, however, identifying the complete VOA is substantially stronger
than matching a central charge or a finite number of character coefficients.
One must distinguish an equality of graded vector spaces from an isomorphism
that also preserves the operator products.

Our first main result is an all-orders character level test of
\eqref{eq:intro-main-correspondence}.  Starting from the conformal gauging
formula, we derive a uniform contour integral expression for the Schur index  $\mathcal{I}_{(A_2,D_{3n-2})}(q)$.
Its gauge dependent part is an exact regrading of the Schur index of
4d $\mathcal N=4$ $SU(2)$ super Yang--Mills theory, while the
remaining factor is the character of a purely bosonic oscillator space.  By
combining this factorization with the known decompositions of the small
$\mathcal N=4$ super Virasoro vacuum module and the doublet algebra
\cite{Buican:2020moo, Feigin:2007sp}, we obtain an exact equivalence of
graded super vector spaces and prove
\begin{equation}
 \mathcal I_{(A_2,D_{3n-2})}(q)
 =
 \operatorname{sch}_{\mathrm{vac}}[\cA(4n-2)](q)
 \label{eq:intro-character-identity}
\end{equation}
for every $n\geq2$.  Here the right hand side is the supercharacter of the vacuum representation of $\cA(4n-2)$.
The same conformal gauging canonically determines a
diagonal $\mathfrak{sl}_2$ BRST complex with the required level and central
charge.  We therefore propose a  BRST cohomology description of  $\cA(4n-2)$.  The
identity \eqref{eq:intro-character-identity} is exact, whereas the full BRST
identification remains conjectural for $n>2$ because the strong generators,
their operator products, and possible cohomology  in nonzero ghost number have
not yet been determined.

The second question concerns the 3d theory obtained by a
$U(1)_r$-twisted circle reduction.  On the three-dimensional side, Costello
and Gaiotto introduced VOAs associated with holomorphic boundary conditions
in topologically twisted 3d $\mathcal N=4$ gauge theories
\cite{Costello:2018fnz}.  For Lagrangian 4d $\mathcal N=2$ conformal gauge
theories, the Schur VOA and the corresponding 3d boundary VOA are naturally
described by the same free-field BRST construction after circle reduction.
This raised the question of how the relation should be formulated for
non-Lagrangian theories such as Argyres--Douglas SCFTs.  Dedushenko developed
a 4d/3d/2d picture in which the 4d theory is placed on a topologically twisted
and Omega-deformed cigar; reduction along its angular direction imposes
$U(1)_r$-twisted boundary conditions, and the 4d Schur VOA is realized as the
boundary VOA of the resulting 3d $\mathcal N=4$ theory
\cite{Dedushenko:2023cvd}.  Related R-twisted compactifications and their 3d
TQFT data had also been studied from BPS monodromy and topological viewpoints
\cite{Cecotti:2010fi, Dedushenko:2018bpp}.

Supersymmetry enhancing 3d $\mathcal N=2$ Chern--Simons (CS) matter theories had
already appeared in \cite{Gang:2018huc}.  The subsequent works
\cite{ArabiArdehali:2024ysy, Gaiotto:2024ioj} made the key
proposal that such theories, equipped with suitable monopole superpotentials,
can provide ultraviolet descriptions flowing to the R-twisted reductions of
4d Argyres--Douglas theories.  The first work derived this bridge from the
high-temperature effective theory of the 4d index, while the second extracted
it from the 4d  BPS spectrum and quantum monodromy.  The proposal
was then extended and tested for the $(A_1,G)$ theories with $G$ of $A$-,
$D$-, and $E$-type, and for more general $(G,G')$ theories, including coprime
$(A_{M-1},A_{N-1})$ families whose associated VOAs are Virasoro or
$W$-minimal models
\cite{ArabiArdehali:2024vli, Gang:2024loa, Go:2025ixu, Kim:2025rog, Nishinaka:2025ytu}.  
Before the $(A_2,D_4)$ construction of \cite{Hamachika:2026whv}, every
example studied in this framework had a trivial Coulomb branch.  That work
provided the first example with a nontrivial Coulomb branch, argued to be
$\mathbb C^2/\mathbb Z_2$.

The present work extends the first nontrivial Coulomb branch example uniformly
to the full family \eqref{eq:intro-AD-family}.  Fermionic character formulas of
Nahm sum type are especially useful for this purpose because the same sums
occur as half-indices of abelian CS matter theories
\cite{ArabiArdehali:2024ysy, Gaiotto:2024ioj}.  The
fermionic formula for $\cA(4n-2)$ therefore supplies a candidate ultraviolet
description for every $n\geq2$, with the $n=2$ construction of
\cite{Hamachika:2026whv} as its first member.

More precisely, the Nahm sum determines a candidate ultraviolet 3d $\mathcal{N}=2$ theory with
gauge group $U(1)^{4n-1}$, CS matrix $K^{(4n-2)}$, and one
charge-one chiral multiplet for each gauge factor.  Its linear exponent fixes
the mixing of the reference R-charge with the topological symmetries.  
We then classify the gauge invariant monopole operators of $R_H=2$ that
preserve two distinguished topological symmetries. This gives a
superpotential with $4n-3$ terms. Within the class considered here, these
are the only gauge invariant monopole terms of $R_H=2$ preserving the two
topological symmetries. The surviving $R_H=0$ monopoles generate
\begin{equation}
 \mathbb C[\mathcal M_C]
 \simeq
 \frac{\mathbb C[X,Y,Z]}{(XY-Z^2)},
 \qquad
 \mathcal M_C\simeq\mathbb C^2/\mathbb Z_2,
 \label{eq:intro-Coulomb ring}
\end{equation}
independently of $n$.  Their topological charges organize $X,Y,Z$ into the
moment-map triplet of an enhanced Coulomb branch $SU(2)$ symmetry and select a
natural candidate for the infrared superconformal R-charge.

The superconformal index provides further protected tests of this
picture.  The use of the index to diagnose infrared supersymmetry enhancement
in 3d CS matter theories follows the strategy of
\cite{Gang:2018huc, Gaiotto:2024ioj, ArabiArdehali:2024vli, Hamachika:2026whv}.  
For arbitrary $n$, we determine its universal expansion through
order $\mathfrak q^{3/2}$.  It contains the Coulomb branch moment maps, the
associated flavor-current contribution, and the pair of extra supercurrent
terms expected for enhancement from $\mathcal N=2$ to $\mathcal N=4$.  The
Higgs and Coulomb limits in the first new examples are consistent with a
trivial Higgs branch and with \eqref{eq:intro-Coulomb ring}.  This universality
does not imply that the complete infrared theories are identical: the $n=2$
and $n=3$ indices already differ at order $\mathfrak q^{9/4}$.  

Finally, we introduce a two variable refinement of the fermionic character.
Using the 3d charge identification together with the $n=2$ Macdonald index
matching, we conjecture that the 4d charge $R+r$ is mapped under the
R-twisted reduction to one of the two 3d charges
$(R_H-R_C\pm J_F)/2$, which are related by the Coulomb branch Weyl
reflection. This leads to a conjectural Macdonald index specialization for
general $n$ and an explicit prediction for $(A_2,D_7)$ through order $q^7$.

The results therefore come with three different levels of certainty.  The
Schur index identities, the graded-vector-space equivalence, the monopole
classification, and the Coulomb ring computation are exact within the stated
constructions.  The identification of the BRST cohomology and the complete
3d infrared flows are proposals supported by protected data.
The Macdonald specialization is a further conjectural refinement motivated by
the $n=2$ example and the uniform 3d charge structure.
Maintaining these distinctions will be important throughout the paper.

\paragraph{Organization.}
Section~\ref{sec:scft-voa} develops the proposed SCFT/VOA correspondence,
derives the Schur index from conformal gauging, proves the all-orders vacuum
supercharacter identity, and formulates the BRST realization.  In
Section~\ref{sec:3d-reduction}, the fermionic character is interpreted as a
3d half-index and the monopole superpotential is determined.
The Coulomb branch chiral ring is determined in
Subsection~\ref{sec:Coulomb chiral-ring}, while
Subsection~\ref{sec:3d-superconformal-index} studies the
superconformal index and the evidence for supersymmetry enhancement.
Section~\ref{sec:Macdonald-refinement} presents the conjectural Macdonald
refinement and the 4d/3d charge map.  We conclude
in Section~\ref{sec:summary-discussion} with a summary and a discussion of the
remaining problems.

\bigskip

\noindent {\bf Note added:} When this paper was nearly completed, Ref. \cite{Jiang:2026keq}  appeared on arXiv. It has some overlap with the content of Section \ref{sec:scft-voa}.

\section{SCFT/VOA correspondence for $(A_2, D_{3n-2})$ AD theory}
\label{sec:scft-voa}

\subsection{Statement of the proposal}
\label{subsec:proposal}

We consider a family of $(A_2,D_{3n-2})$ Argyres--Douglas theories with $n\geq 2$
and propose that their Schur sector chiral algebras are the logarithmic doublet vertex operator (super) algebras $\cA(4n-2)$.
We use the algebraic realization of $\cA(p)$ described in \cite{AdamovicMilas:2013} and the fermionic character formulas of \cite{Feigin:2007sp}.  More precisely, denoting by $\V(\mathcal{T})$ the VOA associated with a 4d $\mathcal{N}=2$ SCFT $\mathcal{T}$, our proposal is
\begin{equation}
 \V\bigl((A_2,D_{3n-2})\bigr)
 \simeq
 \cA(4n-2)
 \label{eq:main-correspondence}
\end{equation}
for all $n\geq 2$.

For the first member, $n=2$, the proposed correspondence is already known.  The theory can be described by diagonally gauging three copies of the $(A_1,A_3)$ Argyres--Douglas theory, and its associated chiral algebra is the $\cA(6)$ algebra \cite{Buican:2016arp}:
\begin{equation}
 \V\bigl((A_2,D_4)\bigr)
 \simeq
 \cA(6).
 \label{eq:n2-known}
\end{equation}
Equation \eqref{eq:main-correspondence} is therefore a proposed extension of this known correspondence to the infinite sequence.

The comparison developed below has four parts.  First, the 2d central charge of $\cA(4n-2)$ agrees exactly with the value dictated by the 4d conformal anomaly.  Second, conformal gauging gives a uniform contour-integral expression for the Schur index of $(A_2,D_{3n-2})$.  
Third, by combining the decomposition of the $c=-9$ small $\mathcal N=4$
super Virasoro algebra in \cite{Creutzig:2018ltv} with the Virasoro module
decomposition of $\cA(4n-2)$, we establish an exact graded vector space
equivalence and prove equality of the vacuum supercharacters to all orders.  
Fourth, the conformal gauging operation suggests a BRST realization of the candidate VOA.  Thus the character level statement is exact for every $n\geq2$, while the identification of the full BRST cohomology with $\cA(4n-2)$ remains a proposal for $n>2$.

\subsection{Central charge matching}
\label{subsec:central-charge}
The 4d/2d correspondence associates the Schur sector of every 4d $\mathcal N=2$ SCFT with a non-unitary vertex operator algebra \cite{Beem:2013sza}.
 The cohomology classes of Schur operators furnish the states of the VOA, and the  Schur index is its vacuum supercharacter. 
 In particular, the 4d stress-tensor multiplet gives rise to the 2d Virasoro field. 
 Comparison of the stress-tensor operator products gives the universal relation
 between the 2d central charge of the VOA and the 4d central charge:
\begin{equation}
 c_{\rm 2d}=-12c_{\rm 4d}.
 \label{eq:4d2d-central-charge-dictionary}
\end{equation}

For the $(G,G')$ Argyres--Douglas theories considered here, the conformal anomaly can be computed from their isolated-hypersurface or irregular-singularity realization and the resulting Coulomb branch data.  The formula of \cite{Wang:2015mra} takes the form
\begin{equation}
 c_{\rm 4d}{(G,G')}
 =
 \frac{r(G)r(G')}{12}
 \left(
  \frac{h(G)h(G')}{h(G)+h(G')}+1
 \right),
 \label{eq:general-c4d}
\end{equation}
where $r(G)$ and $h(G)$ denote the rank and Coxeter number of $G$, respectively. For $G=A_2$ and $G'=D_m$,
\begin{equation}
 r(A_2)=2,
 \qquad h(A_2)=3,
 \qquad r(D_m)=m,
 \qquad h(D_m)=2m-2.
\end{equation}
Substitution into \eqref{eq:general-c4d} gives
\begin{equation}
 c_{\rm 4d}{(A_2,D_m)}
 =
 \frac{m(8m-5)}{6(2m+1)}.
 \label{eq:c4d-A2Dm}
\end{equation}
Setting $m=3n-2$, we obtain
\begin{equation}
 c_{\rm 4d} {(A_2,D_{3n-2})}
 =
 \frac{(3n-2)(8n-7)}{6(2n-1)}.
 \label{eq:c4d-family}
\end{equation}

The central charge of the $\cA(p)$ algebra is \cite{Feigin:2007sp}
\begin{equation}
 c_{\rm 2d}({\cA(p)})
 =
 13-6p-\frac{6}{p}.
 \label{eq:c-Ap}
\end{equation}
For $p=4n-2$, this becomes
\begin{equation}
 c_{\rm 2d}({\cA(4n-2)})
 =
 25-24n-\frac{3}{2n-1}.
 \label{eq:c-A4n2}
\end{equation}
Using \eqref{eq:c4d-family}, one finds the exact relation
\begin{equation}
 c_{\rm 2d}({\cA(4n-2)})
 =
 -12c_{\rm 4d}{(A_2,D_{3n-2})},
 \label{eq:central-charge-match}
\end{equation}
which is precisely the dictionary \eqref{eq:4d2d-central-charge-dictionary}.  This equality is a necessary consistency condition and, in particular, singles out $p=4n-2$ among the integer parameters $p\geq2$ of the doublet family.  It is not by itself an identification of VOAs, since the central charge does not determine the operator content or the operator products.  The all-orders character identity and the BRST construction developed below provide substantially stronger tests of \eqref{eq:main-correspondence}.

\subsection{Schur index from conformal gauging}
\label{subsec:schur-gauging}

The conformal gauging description used below corresponds to a weakly coupled cusp of the conformal manifold. Such weakly coupled descriptions arise at degeneration limits in which strongly coupled sectors are connected through a weakly gauged common symmetry.  This degeneration was analyzed systematically for the $(A_m,D_k)$ and $D^b_p(SO(2N))$ families in \cite{Carta:2021whq}, where the frame relevant here was identified.  In this frame the $(A_2,D_{3n-2})$ theory is represented by diagonally gauging the $SO(3)$ flavor symmetries of two Argyres--Douglas sectors:
\begin{equation}
 (A_2,D_{3n-2})
 =
 \bigl[
 D^2_{2n-1}(SO(3))
 \longleftarrow SO(3) \longrightarrow
 D_{2n-1}^{2n-2}\bigl(SO(2n),[2n-3,1^3]\bigr)
 \bigr].
 \label{eq:conformal gauging}
\end{equation}
Note that the Schur index and BRST computations below depend only on the Lie algebra $\mathfrak{so}(3)\simeq\mathfrak{su}(2)$
and are insensitive to this choice of global form.
The first sector is equivalent to
\begin{equation}
 D^2_{2n-1}(SO(3))
 \simeq
 (A_1,D_{2n-1}).
\end{equation}
The gauge coupling parametrizes an exactly marginal direction.  Equivalently, the matter contributions to the $SO(3)$ beta function cancel the vector multiplet contribution. The 2d counterpart of the same cancellation is the nilpotency condition for the BRST charge used in Subsection~\ref{subsec:BRST}.

The Schur index is independent of exactly marginal couplings, so it may be evaluated at this weakly coupled cusp. 
The  gauging prescription multiplies the Schur indices of the two matter sectors by the vector multiplet contribution and integrates over the gauge holonomy.     This gives the contour integral formula for the Schur index 
\begin{align}
 \cI_{(A_2,D_{3n-2})}(q)
 &=
 \frac{1}{2}
 \oint \frac{dz}{2\pi i z}
 (1-z^2)(1-z^{-2})
 \cI_{\mathrm{vec}}^{SO(3)}(z,q)
 \nonumber\\
 &\hspace{2.0em}\times
 \cI_{D^2_{2n-1}(SO(3))}(z,q)
 \cI_{D_{2n-1}^{2n-2}(SO(2n),[2n-3,1^3])}(z,q).
 \label{eq:schur-gauging-integral}
\end{align}
Here $dz/(2\pi iz)$ together with $(1-z^2)(1-z^{-2})$ is the Haar measure of $SO(3)$.  Thus the contour extracts precisely the gauge singlet operators. The vector multiplet contribution and the first Argyres--Douglas block are \cite{Song:2017oew}
\begin{align}
 \cI_{\mathrm{vec}}^{SO(3)}(z,q)
 &=
 \PE\left[
 -\frac{2q}{1-q}\chi_{\mathbf{3}}(z)
 \right],
 \label{eq:vector-index}
 \\
 \cI_{D^2_{2n-1}(SO(3))}(z,q)
 &=
 \PE\left[
 \frac{q-q^{2n-1}}
 {(1-q)(1-q^{2n-1})}
 \chi_{\mathbf{3}}(z)
 \right].
 \label{eq:first-block-index}
\end{align}
Here $\chi_{\mathbf{3}}(z)$ is the character of the adjoint  representation of $SU(2)$:  
\begin{equation}
 \chi_{\mathbf{3}}(z)
 =
 z^{-2}+1+z^2
 \label{eq:chi3}
\end{equation}
and $\PE[\cdots]$ is the plethystic exponential defined by   
\begin{equation}
 \PE[f(x_1,x_2,\ldots)]
 :=
 \exp\left(
 \sum_{k=1}^{\infty}
 \frac{1}{k}f(x_1^k,x_2^k,\ldots)
 \right).
 \label{eq:PE-definition}
\end{equation}

The $D^b_p(G,Y)$ theories may be constructed by compactifying the six-dimensional $(2,0)$ theory on a sphere with an irregular puncture and a regular puncture.  The partition $Y$ labels a partial closure of the regular puncture.  In 4d language this is implemented by giving the moment map operator a nilpotent vacuum expectation value, whereas in the associated VOA description it is encoded by quantum Drinfeld--Sokolov reduction.  This relation between nilpotent Higgsing, puncture closure, and $W$-algebras underlies the index formula derived in \cite{Song:2017oew}.

In general, $D^b_p(G,Y)$ denotes the theory obtained from $D^b_p(G)$ by the nilpotent deformation labelled by $Y$. 
The unbroken flavor symmetry group $G_F$ is the commutant of the embedding
\begin{equation}
 \rho_Y:SU(2)\longrightarrow G\,,
\end{equation}
where $\rho_Y$ is the $SU(2)$ embedding associated with the nilpotent orbit labelled by the partition $Y$.
For $b=h(G)$, the TQFT wave function construction of the Schur index gives \cite{Song:2017oew}
\begin{equation}
 \cI_{D^h_p(G,Y)}(z,q)
 =
 \PE\left[
 \sum_j\frac{q^{j+1}}{1-q}\chi_{R_j}(z)
 -
 \frac{q^p}{1-q^p}
 \chi_{\adj}^{\rho_Y}(z,q)
 \right].
 \label{eq:Dhp-index}
\end{equation}
The representations $R_j$ of the residual flavor algebra $\mathfrak{g}_F$ are defined by the decomposition
\begin{equation}
 \adj_{\mathfrak{g}}
 =
 \bigoplus_j V_j\otimes R_j
 \label{eq:adjoint-decomposition-general}
\end{equation}
under $\rho_Y(\su_2)\oplus\mathfrak{g}_F\subset\mathfrak{g}$, where $V_j$ is the spin-$j$ representation of $\su_2$. Correspondingly,
\begin{equation}
 \chi_{\adj}^{\rho_Y}(z,q)
 =
 \sum_j \chi_{V_j}(q)\chi_{R_j}(z),
 \qquad
 \chi_{V_j}(q)=\sum_{s=-j}^{j}q^s\,.
 \label{eq:rho-character}
\end{equation}
The first term in \eqref{eq:Dhp-index} records the fields associated with the $\su_2$ multiplets that survive the partial closure, organized by their residual-flavor representations $R_j$.  The second term is the irregular puncture contribution evaluated after the specialization determined by $\rho_Y$.  
The same expression agrees with the expected vacuum character of the corresponding W-algebra, as expected from the correspondence between the 4d nilpotent vev deformation and the 2d Drinfeld--Sokolov reduction.
To evaluate \eqref{eq:Dhp-index} for the $D_{2n-1}^{2n-2}(SO(2n),[2n-3,1^3])$ theory, it remains to determine the decomposition of the $SO(2n)$ adjoint under the embedding associated with $Y=[2n-3,1^3]$.  We carry out this step next.

We now specialize \eqref{eq:Dhp-index} to
\begin{equation}
 G=SO(2n),
 \qquad
 Y=[2n-3,1^3].
\end{equation}
For $n\geq3$, the connected commutant of the corresponding nilpotent
embedding is $G_F=SO(3)$.  At $n=2$, the orbit $Y=[1^4]$ is trivial and
the full commutant is $SO(4)$; in the formulas below, $z$ keeps track of
the diagonal subgroup $SO(3)\subset SO(4)$ that participates in the
conformal gauging.  Under this subgroup, the vector representation obeys
$\mathbf{4}=\mathbf{1}\oplus\mathbf{3}$, so the following formulas apply
uniformly for all $n\geq2$.  Empty sums are understood to vanish at $n=2$.

The vector representation decomposes as
\begin{equation}
 \mathbf{2n}
 =
 \bigl(V_{n-2}\otimes\mathbf{1}\bigr)
 \oplus
 \bigl(V_0\otimes\mathbf{3}\bigr).
 \label{eq:vector-decomposition}
\end{equation}
Since $\adj_{\so_{2n}}=\wedge^2\mathbf{2n}$, we use
\begin{equation}
 \wedge^2 V_{n-2}
 =
 \bigoplus_{r=0}^{n-3}V_{2r+1}
 \label{eq:wedge-spin}
\end{equation}
to obtain
\begin{equation}
 \adj_{\so_{2n}}
 =
 \bigl(V_0\otimes\mathbf{3}\bigr)
 \oplus
 \bigl(V_{n-2}\otimes\mathbf{3}\bigr)
 \oplus
 \bigoplus_{r=0}^{n-3}
 \bigl(V_{2r+1}\otimes\mathbf{1}\bigr).
 \label{eq:adjoint-decomposition-so2n}
\end{equation}
Thus the  character is
\begin{equation}
 \chi_{\adj_{\so_{2n}}}^{\rho_Y}(z,q)
 =
 \sum_{r=0}^{n-3}\chi_{V_{2r+1}}(q)
 +
 \bigl(1+\chi_{V_{n-2}}(q)\bigr)\chi_{\mathbf{3}}(z).
 \label{eq:adjoint-character-so2n}
\end{equation}
The first term in the exponent of \eqref{eq:Dhp-index} is determined from the same decomposition:
\begin{equation}
 \sum_j\frac{q^{j+1}}{1-q}\chi_{R_j}(z)
 =
 \frac{1}{1-q}
 \left[
 \sum_{r=0}^{n-3}q^{2r+2}
 +
 \bigl(q+q^{n-1}\bigr)\chi_{\mathbf{3}}(z)
 \right].
 \label{eq:positive-letter-index}
\end{equation}
Substituting \eqref{eq:adjoint-character-so2n} and \eqref{eq:positive-letter-index} into \eqref{eq:Dhp-index}, we find
\begin{equation}
 \cI_{D^{2n-2}_{2n-1}(SO(2n),[2n-3,1^3])}(z,q)
 =
 \PE\left[
 f_{D_n}(z,q)
 \right],
 \label{eq:second-block-PE}
\end{equation}
where
\begin{align}
 f_{D_n}(z,q)
 &=
 \frac{1}{1-q}
 \left[
 \sum_{r=0}^{n-3}q^{2r+2}
 +
 \bigl(q+q^{n-1}\bigr)\chi_{\mathbf{3}}(z)
 \right]
 \nonumber\\
 &\quad-
 \frac{q^{2n-1}}{1-q^{2n-1}}
 \left[
 \sum_{r=0}^{n-3}\chi_{V_{2r+1}}(q)
 +
 \bigl(1+\chi_{V_{n-2}}(q)\bigr)\chi_{\mathbf{3}}(z)
 \right]
 \label{eq:fDn-first}
 \\
 &=
 \frac{1}{1-q^{2n-1}}
 \left[
 \frac{q^2(1-q^{2n-4})}{1-q}
 +
 \left
 \{
 \frac{q(1-q^{2n-2})}{1-q}
 +q^{n-1}(1+q)
 \right\}
 \chi_{\mathbf{3}}(z)
 \right].
 \label{eq:fDn-simplified}
\end{align}

Combining \eqref{eq:vector-index}, \eqref{eq:first-block-index}, and \eqref{eq:fDn-simplified}, the Schur index of the full theory can be written in the compact form
\begin{equation}
 \cI_{(A_2,D_{3n-2})}(q)
 =
 \frac{1}{2}
 \oint\frac{dz}{2\pi i z}
 (1-z^2)(1-z^{-2})
 \PE\left[f_n(z,q)\right],
 \label{eq:full-schur-compact}
\end{equation}
with
\begin{equation}
 f_n(z,q)
 =
 \frac{q^2}{1-q^{2n-1}}
 \left[
 \frac{1-q^{2n-4}}{1-q}
 +
 \left(
 q^{n-3}(1+q)-2q^{2n-3}
 \right)
 \chi_{\mathbf{3}}(z)
 \right].
 \label{eq:full-single-letter}
\end{equation}
Equation \eqref{eq:full-schur-compact} is the general Schur index formula that follows from the conformal gauging description \eqref{eq:conformal gauging}.

The compact formula also admits an exact rewriting in terms of the Schur index of 4d $\mathcal N=4$ $SU(2)$ super Yang--Mills theory.   The single letter index for 4d $\mathcal N=4$ $SU(2)$ super Yang--Mills theory is 
\begin{equation}
 f_{\mathrm{SYM}}(z,x,q)
 =
 \frac{q^{1/2}(x+x^{-1})-2q}{1-q}
 \chi_{\mathbf{3}}(z)\,,
 \label{eq:N4-SYM-letter}
\end{equation}
where $x$ is the $U(1)$ flavor fugacity for the adjoint hypermultiplet.
Then the full single letter index \eqref{eq:full-single-letter} decomposes exactly as
\begin{align}
 f_n(z,q)
 &=s_n(q)+f_{\mathrm{SYM}}\bigl(z,q^{1/2},q^{2n-1}\bigr),
 \label{eq:fn-SYM-decomposition}
 \\
 s_n(q)
 &:=\frac{q^2(1-q^{2n-4})}{(1-q)(1-q^{2n-1})}.
 \label{eq:singlet-factor-sn}
\end{align}
Since $s_n(q)$ is independent of the gauge fugacity, it factors out of the contour integral.  If $\cI_{\mathrm{SYM}}(q;x)$ 
denotes the flavored Schur index for 
the $\mathcal N=4$ $SU(2)$ super Yang--Mills theory, we therefore obtain
\begin{equation}
 \cI_{(A_2,D_{3n-2})}(q)
 =
 \PE\bigl[s_n(q)\bigr]\,
 \cI_{\mathrm{SYM}}\bigl(q^{2n-1};q^{1/2}\bigr).
 \label{eq:AD-SYM-factorization}
\end{equation}

For $n=2$, the singlet factor vanishes, $s_2(q)=0$, and \eqref{eq:AD-SYM-factorization} reduces to
\begin{equation}
 \cI_{(A_2,D_4)}(q)
 =
 \cI_{\mathrm{SYM}}(q^3;q^{1/2}).
 \label{eq:A6-SYM-index-relation}
\end{equation}
This is precisely the index relation between the $(A_2, D_4)$ Argyres--Douglas theory and $\mathcal N=4$ $SU(2)$ super Yang--Mills theory obtained in \cite{Buican:2020moo}.

\subsection{Graded vector space isomorphism and the Schur index identity}
\label{subsec:generalized-GVSI}
Let $\mathfrak S$ denote the small $\mathcal N=4$ super-Virasoro algebra at
$c=-9$. It contains the affine vertex algebra
$V_{-3/2}(\mathfrak{sl}_2)$ and admits a commuting $\mathfrak{sl}_2$ action. As shown in \cite{Creutzig:2018ltv}, it decomposes as
\begin{equation}
\mathfrak S
\simeq
\bigoplus_{m=0}^{\infty}
\pi_{m+1}\otimes V_{-3/2}(m\omega),
\end{equation}
where $V_{-3/2}(m\omega)$ is the Weyl module of  $V_{-3/2}$, while  $\pi_{m+1}$ is the $(m+1)$-dimensional irreducible representation of  the commuting $\mathfrak{sl}_2$ action. On the other hand, the doublet algebra has the exact decomposition \cite{Feigin:2007sp}
\begin{equation}
\cA(p)
\simeq
\bigoplus_{m=0}^{\infty}
\pi_{m+1}\otimes M_{m+1,1;p},
\end{equation}
where $M_{m+1,1;p}$ is the irreducible Virasoro module with Kac labels $(m+1,1)$.

For $p=4n-2$, introduce the bosonic graded vector space
\begin{equation}
 \mathcal F_n
 :=
 \operatorname{Sym}
 \left(
 \bigoplus_{\ell\geq0}
 \bigoplus_{r=2}^{2n-3}
 \C b_{(2n-1)\ell+r}
 \right),
 \qquad
 \deg b_j=j.
 \label{eq:Fn-definition}
\end{equation}
Here $ \operatorname{Sym}(V):=\bigoplus_{k \ge 0}\operatorname{Sym}^k(V)$. 
The inner sum is empty for $n=2$, so that $\mathcal F_2=\C$.  Its character is
\begin{align}
 \operatorname{ch}\mathcal F_n
 &=
 \prod_{\ell\geq0}
 \prod_{r=2}^{2n-3}
 \frac{1}{1-q^{(2n-1)\ell+r}}
 \nonumber\\
 &=
 \PE\left[
 \frac{q^2+q^3+\cdots+q^{2n-3}}
 {1-q^{2n-1}}
 \right]
 =
 \PE[s_n(q)].
 \label{eq:Fn-character}
\end{align}

We now introduce a new grading on the small $\mathcal N=4$ algebra. Let $h$ denote the conformal weight, namely the $L_0$ eigenvalue. The zero modes $J^0_0$ and $J^\pm_0$ of  the affine $\slc_2$ current generate an ordinary $\slc_2$ algebra, and 
we denote its Cartan weight by $f$.
Under the 4d/2d correspondence, this $\mathfrak{sl}_2$ symmetry is associated with  the $SU(2)_F$ flavor symmetry of $\mathcal N=4$ $SU(2)$ super Yang--Mills theory. We define
\begin{equation}
\deg_n=(2n-1)h+\frac12 f.
\label{eq:generalized-GVSI-degree}
\end{equation}
The highest weight state of $V_{-3/2}(m\omega)$ has Sugawara conformal weight
\begin{equation}
 h_m=\frac{m(m+2)}{2}.
 \label{eq:affine-highest-weight-hm}
\end{equation}
It forms an $(m+1)$-dimensional representation under the zero-mode $\mathfrak{sl}_2$. Its states may be labelled by $s=0,1,\ldots,m$, with Cartan weights
\begin{equation}
 f_s=2s-m.
 \label{eq:horizontal-flavor-weights}
\end{equation}
Their regraded degrees are therefore
\begin{equation}
 (2n-1)h_m+\frac12 f_s
 =
 \Delta_m^{(n)}+s,
 \qquad
 \Delta_m^{(n)}
 :=
 (2n-1)\frac{m(m+2)}{2}-\frac m2.
 \label{eq:horizontal-regraded-degrees}
\end{equation}
The affine current modes  $J^-_{-k}$, $J^0_{-k}$, and $J^+_{-k}$ 
have Cartan weights $-2, 0,$ and $2$, respectively.  
Therefore, under the grading defined above, their degrees are
\begin{align}
    (2n-1)k-1\,, \quad (2n-1)k\,, \quad (2n-1)k+1\,.
\end{align}
Together with the zero-mode $\mathfrak{sl}_2$ multiplet, whose states
have degrees
\begin{align}
\Delta_m^{(n)}\,, \Delta_m^{(n)}+1,\ldots,\Delta_m^{(n)}+m\,,
\end{align}
this gives
\begin{align}
 \operatorname{ch}_{\deg_n}V_{-3/2}(m\omega)
 ={}&q^{\Delta_m^{(n)}}
 (1+q+\cdots+q^m)
 \prod_{k=1}^{\infty}
 \frac{1}
 {(1-q^{(2n-1) k-1})(1-q^{(2n-1)k})(1-q^{(2n-1) k+1})}\,.
 \label{eq:Weyl-regraded-character}
 \end{align}
The generators of $\mathcal F_n$ have degrees  $2,3,\ldots,2n-3$ modulo $2n-1$, whereas the affine currents have degrees   $-1,0,1$ modulo $2n-1$.
Together, these account for exactly one bosonic oscillator at every degree
$j\geq2$. Hence, using \eqref{eq:Weyl-regraded-character}, we obtain
\begin{equation}
 \operatorname{ch}\mathcal F_n\,
 \operatorname{ch}_{\deg_n}V_{-3/2}(m\omega)
 =
 \frac{q^{\Delta_m^{(n)}}(1-q^{m+1})}{(q;q)_\infty}.
 \label{eq:Fn-Weyl-character}
\end{equation}
We compare this expression with the corresponding Virasoro module.
For $p=4n-2$, the conformal weight of the highest weight state in 
$M_{m+1,1;p}$ is 
\begin{align}
 h_{m+1,1}^{(1,4n-2)}
 &=
 (2n-1)\frac{m(m+2)}{2}-\frac{m}{2}
 =
 \Delta_m^{(n)},
 \label{eq:Virasoro-Kac-weight-general}
\end{align}
Moreover, $M_{m+1,1;4n-2}$ is obtained from its Verma module by
quotienting by the submodule generated by a singular vector at level $m+1$ \cite{Feigin:2007sp}. Its character is therefore
\begin{equation}
 \operatorname{ch}M_{m+1,1;4n-2}
 =
 \frac{q^{\Delta_m^{(n)}}(1-q^{m+1})}{(q;q)_\infty}.
 \label{eq:Virasoro-module-character-general}
\end{equation}
Since every graded component is finite-dimensional, \eqref{eq:Fn-Weyl-character} and \eqref{eq:Virasoro-module-character-general} give a noncanonical linear equivalence of graded vector spaces for every $m$,
\begin{equation}
 \mathcal F_n\otimes V_{-3/2}(m\omega)
 \simeq_{\mathrm{gr}}
 M_{m+1,1;4n-2}.
 \label{eq:module-generalized-GVSI}
\end{equation}
For $p=4n-2$, the doublet generators of $\cA(p)$ are odd \cite{AdamovicMilas:2013}.  The $m$th summand in the doublet decomposition is generated from the vacuum sector by an $m$-fold product of these odd generators and therefore has parity $(-1)^m$.  The corresponding $m$-th summand in the small $\mathcal N=4$ decomposition has the same parity, as in the $n=2$ construction of \cite{Buican:2020moo}.  Combining the decomposition of $\mathfrak{S}$ and $\mathcal{A}(4n-2)$ with 
the graded isomorphism above, we obtain
\begin{equation}
 \mathcal F_n\otimes\mathfrak S
 \simeq_{\mathrm{gr}}
 \cA(4n-2)
 \label{eq:generalized-GVSI}
\end{equation}
as $\mathbb Z_{\geq0}\times\mathbb Z_2$-graded vector spaces. The grading on
$\mathfrak S$ is given by
\begin{equation}
 \deg_n=(2n-1)h+\frac12 f.
\end{equation}
For  $n=2$, $\mathcal F_2=\mathbb{C}$ and this relation reduces to the graded vector space isomorphism established in \cite{Buican:2020moo}.

Taking the supercharacter of \eqref{eq:generalized-GVSI} gives
\begin{align}
 \operatorname{sch}_{\mathrm{vac}}[\cA(4n-2)](q)
 &=
 \PE[s_n(q)]\,
 \cI_{\mathrm{SYM}}(q^{2n-1};q^{1/2})
 \nonumber\\
 &=
 \frac{1}{(q;q)_\infty}
 \sum_{m=0}^{\infty}
 (-1)^m(m+1)
 q^{\Delta_m^{(n)}}(1-q^{m+1}).
 \label{eq:Ap-general-supercharacter}
\end{align}
Together with the Schur index formula \eqref{eq:AD-SYM-factorization}, this proves the all orders identity
\begin{equation}
 \cI_{(A_2,D_{3n-2})}(q)
 =
 \operatorname{sch}_{\mathrm{vac}}[\cA(4n-2)](q)
 \label{eq:all-orders-index-character}
\end{equation}
for every $n\geq2$.

Equation \eqref{eq:generalized-GVSI} is an exact statement about graded super vector spaces.  We do not construct a map preserving operator products, nor do we establish the additional compatibility with normal ordered products and derivatives proved at $n=2$ in \cite{Buican:2020moo}.  It therefore does not by itself imply an isomorphism of VOAs and should be distinguished from the BRST cohomology proposal discussed below.

\subsection{$\cA(4n-2)$ as a BRST cohomology}
\label{subsec:BRST}

The diagonal conformal gauging in four dimensions has a direct counterpart in the VOA: one tensors the VOAs of the two matter sectors with an adjoint $bc$ ghost system and takes the BRST cohomology with respect to the diagonally gauged affine $\slc_2$ symmetry \cite{Beem:2013sza,Buican:2016arp}.  We write $L_k(\mathfrak g)$ and $W_k(\mathfrak g,f)$ for the simple quotients of the corresponding universal affine and $W$-algebras.  Let
\begin{align}
 \V_1
 &: =
 L_{k_1}(\slc_2),
 &
 k_1
 &: =
 -2+\frac{2}{2n-1},
 \label{eq:V1-level}
 \\
 \V_2
 &: =
 W_{k_{\so}}
 \bigl(\so_{2n},f_{[2n-3,1^3]}\bigr),
 &
 k_{\so}
 &: =
 -\frac{(2n-2)^2}{2n-1}.
 \label{eq:V2-level}
\end{align}
Here $\V_1$ is the affine VOA associated with the first matter sector in \eqref{eq:conformal gauging}, while $\V_2$ is the simple quantum Drinfeld--Sokolov reduction associated with the nilpotent orbit $[2n-3,1^3]$ in the second sector.  The subscript $k_{\so}$ in \eqref{eq:V2-level} denotes the level of the affine $\so_{2n}$ algebra before Drinfeld--Sokolov reduction.  The reduction leaves an affine $\slc_2$ current algebra corresponding to the unbroken $SO(3)$ flavor symmetry of the second matter sector.

Denote the affine $\mathfrak{sl}_2$ currents in $\V_1$ and $\V_2$ by $J^a_{(1)}$ and $J^a_{(2)}$, and let $k_2$ be the level of 
the residual affine $\mathfrak{sl}_2$ algebra in $\V_2$. 
Under the 4d/2d correspondence, the 4d conformal gauging condition requires 
\begin{align}
    k_1+k_2 = -2 h^{\vee}(A_1)=-4\,.
 \label{eq:BRST-level-condition}
\end{align}
Here $h^{\vee}(A_1)$ is the dual Coxeter number of $A_1$.
Since
\begin{equation}
 k_1
 =
 -2+\frac{2}{2n-1}\,,
\end{equation}
this determines
\begin{equation}
 k_2
 =
 -2-\frac{2}{2n-1}.
 \label{eq:second-sl2-level}
\end{equation}
Thus the condition for vanishing of the 4d beta function is 
mapped to the level condition required for the BRST reduction of the diagonal affine $\mathfrak{sl}_2$.

Introduce adjoint ghosts $b_a$ and $c^a$ of conformal weights one and zero, respectively, with operator product
\begin{equation}
 b_a(z)c^b(w)
 \sim
 \frac{\delta_a{}^b}{z-w}.
 \label{eq:bc-OPE}
\end{equation}
For a basis $T_a$ satisfying $[T_a,T_b]=f_{ab}{}^cT_c$, the BRST current and charge are
\begin{align}
 j_{\mathrm{BRST}}(z)
 &=
 c^a(z)\bigl(J_{(1),a}(z)+J_{(2),a}(z)\bigr)
 -\frac{1}{2}f_{ab}{}^c
 :c^a(z)c^b(z)b_c(z):,
 \label{eq:BRST-current}
 \\
 Q_{\mathrm{BRST}}
 &=
 \oint\frac{dz}{2\pi i}\,j_{\mathrm{BRST}}(z).
 \label{eq:BRST-charge}
\end{align}
Equation \eqref{eq:BRST-level-condition} implies $Q_{\mathrm{BRST}}^2=0$.  The VOA obtained  by the gauging is therefore the BRST cohomology
\begin{equation}
 \V_{\mathrm{gauge}}
 :=
 H^\bullet_{\mathrm{BRST}}\!\left(
 \V_1\otimes\V_2\otimes bc_{\slc_2},
 Q_{\mathrm{BRST}}
 \right).
 \label{eq:gauged-VOA}
\end{equation}
The proposal \eqref{eq:main-correspondence} can now be stated more precisely as the isomorphism
\begin{equation}
 \cA(4n-2)
 \simeq
 \V_{\mathrm{gauge}}
\,.
 \label{eq:BRST-Ap}
\end{equation}

The central charge gives a useful internal check of this construction.  The first affine factor and the ghost system contribute
\begin{equation}
 c_{\rm 2d}(\V_1)
 =
 \frac{3k_1}{k_1+2}
 =
 -6(n-1),
 \qquad
 c_{\rm 2d}\bigl(bc_{\slc_2}\bigr)
 =
 -2\dim\slc_2
 =
 -6.
 \label{eq:BRST-central-components}
\end{equation}
For the orbit $[2n-3,1^3]$, the Drinfeld--Sokolov central charge of the second factor is
\begin{equation}
 c_{\rm 2d}(\V_2)
 =
 25-18n-\frac{3}{2n-1}.
 \label{eq:V2-central-charge}
\end{equation}
It follows that the total central charge of the complex is
\begin{equation}
 c_{\rm 2d}(\V_1)+c_{\rm 2d}(\V_2)+c_{\rm 2d}\bigl(bc_{\slc_2}\bigr)
 =
 25-24n-\frac{3}{2n-1}
 =
 c_{\rm 2d}({\cA(4n-2)}),
 \label{eq:BRST-central-match}
\end{equation}
consistent with \eqref{eq:central-charge-match}.

For $n=2$, the orbit $[2n-3,1^3]=[1^4]$ is trivial and $\so_4\simeq\slc_2\oplus\slc_2$.  The second factor then reduces to two copies of $L_{-4/3}(\slc_2)$, while the first factor is another copy at the same level.  Thus \eqref{eq:BRST-Ap} becomes the relative diagonal BRST reduction of three copies of $L_{-4/3}(\slc_2)$.  By the chiral-algebra gauging prescription and the known $n=2$ correspondence, this reduction is expected to reproduce $\cA(6)$ \cite{Buican:2016arp}.

For $n>2$, equation \eqref{eq:BRST-Ap} is a proposed identification rather than a proven isomorphism.  The conformal gauging construction canonically determines the BRST complex, and the level and central charge checks above show that it has the required consistency properties.  The all orders supercharacter identity \eqref{eq:all-orders-index-character} supplies a further necessary check.  It does not, however, prove that the cohomology is concentrated in the expected degree or identify the resulting operator products.  A proof of \eqref{eq:BRST-Ap} would still require, for example, an identification of strong generators and their OPEs, or a direct computation of the BRST cohomology.

\section{The 3d R-twisted reduction}
\label{sec:3d-reduction}

\subsection{The ultraviolet Chern--Simons matter theory}
\label{subsec:UV-CSM-theory}
The vacuum supercharacter of the logarithmic vertex algebra $\cA(p)$ admits
the following Nahm sum expression \cite{Feigin:2007sp}:
\begin{equation}
 \operatorname{sch}_{\mathrm{vac}}[\cA(p)](q)
 =
 \sum_{\bm \ell\in\Z_{\geq0}^{p+1}}
 \frac{
 q^{\frac12\bm \ell^{\mathsf T}K^{(p)}\bm \ell}
 \bigl(-q^{1/2}\bigr)^{B^{(p)}\cdot\bm \ell}
 }{
 \displaystyle\prod_{a=1}^{p+1}(q;q)_{\ell_a}
 },
 \label{eq:Ap-Nahm-sum}
\end{equation}
where
\begin{equation}
 (q;q)_\ell:=\prod_{j=1}^{\ell}(1-q^j),
 \qquad
 (q;q)_0:=1.
 \label{eq:q-Pochhammer}
\end{equation}
The symmetric $(p+1)\times(p+1)$ matrix $K^{(p)}$ is
\begin{equation}
 K^{(p)}=
 \left(
 \begin{array}{ccccccc}
 \frac p2 & \frac p2 & 1 & 2 & 3 & \cdots & p-1\\
 \frac p2 & \frac p2 & 1 & 2 & 3 & \cdots & p-1\\
 1&1&2&2&2&\cdots&2\\
 2&2&2&4&4&\cdots&4\\
 3&3&2&4&6&\cdots&6\\
 \vdots&\vdots&\vdots&\vdots&\vdots&\ddots&\vdots\\
 p-1&p-1&2&4&6&\cdots&2(p-1)
 \end{array}
 \right)\,.
 \label{eq:Kp-matrix}
\end{equation}
Equivalently, the entries of the matrix are
\begin{align}
 K^{(p)}_{ij}
 &=\frac p2,
 &&i,j\in\{1,2\},
 \nonumber\\
 K^{(p)}_{i,a}=K^{(p)}_{a,i}
 &=a-2,
 &&i\in\{1,2\},\quad 3\leq a\leq p+1,
 \nonumber\\
 K^{(p)}_{ab}
 &=2\min(a-2,b-2),
 &&3\leq a,b\leq p+1.
 \label{eq:Kp-components}
\end{align}
This component form fixes the continuation represented by the ellipses in \eqref{eq:Kp-matrix}.
The linear term is given by
\begin{equation}
 B^{(p)}=
 \bigl(p-1,\ p-1,\ 2,\ 4,\ 6,\ldots,\ 2(p-1)\bigr).
 \label{eq:Bp-vector}
\end{equation}

The all orders identity \eqref{eq:all-orders-index-character}  gives the exact Nahm sum representation
\begin{equation}
 \cI_{(A_2,D_{3n-2})}(q)
 =
 \sum_{\bm \ell\in\Z_{\geq0}^{4n-1}}
 \frac{
 q^{\frac12\bm \ell^{\mathsf T}K^{(4n-2)}\bm \ell}
 \bigl(-q^{1/2}\bigr)^{B^{(4n-2)}\cdot\bm \ell}
 }{
 \displaystyle\prod_{a=1}^{4n-1}(q;q)_{\ell_a}
 }
 \label{eq:AD-Nahm-sum}
\end{equation}
for the Schur index of $(A_2,D_{3n-2})$ Argyres--Douglas theory.

The form of \eqref{eq:AD-Nahm-sum} suggests a direct 3d
interpretation, following the relation between Nahm sums and half-indices of
abelian CS matter theories.
Consider a 3d $\mathcal N=2$ abelian CS matter theory
with gauge group
\begin{equation}
 G_{\mathrm{3d}}=U(1)^{4n-1},
 \label{eq:3d-gauge-group}
\end{equation}
whose effective gauge CS level matrix, in the half-index convention, is $K^{(4n-2)}$.  For each factor $U(1)_a$, introduce one chiral multiplet of charge one under $U(1)_a$ and neutral under the remaining gauge factors.  
Supersymmetric  indices of 3d $\mathcal N=2$ theories on $S^1\times D^2$ with 2d $\mathcal N=(0,2)$ boundary supersymmetry
\begin{equation}
 I\!\!I(q, \boldsymbol{x})
 =
 \operatorname{Tr}_{\mathcal H_{D^2}}
 (-1)^F q^{j+R/2}
 \prod_{a} x_a^{J_a},
 \label{eq:half-index-definition}
\end{equation}
were studied in \cite{Beem:2012mb, Yoshida:2014ssa, Dimofte:2017tpi}.  
Here $j$ is a generator of  rotations of the hemisphere $D^2$, $R$ is an   R-charge, and $J_a$ is a generator of a $U(1)$ global symmetry.
For the Dirichlet boundary condition $\mathcal D$ for the vector multiplets and the deformed Dirichlet  boundary condition $D_c$ for the chiral multiplets \cite{Dimofte:2017tpi},  the half-index is 
\begin{equation}
 I\!\!I_{(\mathcal D,D_c)}(q,\bm x)
 =
 \sum_{\bm \ell\in\Z_{\geq0}^{4n-1}}
 \frac{
 q^{\frac12\bm \ell^{\mathsf T}K^{(4n-2)}\bm \ell}
 }{
 \displaystyle\prod_{a=1}^{4n-1}(q;q)_{\ell_a}
 }
 \prod_{a=1}^{4n-1}x_a^{-\ell_a}.
 \label{eq:abelian-half-index}
\end{equation}
Here $x_a$ is the fugacity for the topological symmetry $U(1)_{J_a}$
associated with the gauge factor $U(1)_a$. We follow the boundary magnetic
flux convention of \cite{Dimofte:2017tpi}, in which the sector labelled by
$\bm \ell$ is weighted by $\prod_a x_a^{-\ell_a}$.

The specialization
\begin{equation}
 x_a
 =
 \bigl(-q^{-1/2}\bigr)^{B^{(4n-2)}_a}
,
 \qquad
 a=1,\ldots,4n-1,
 \label{eq:Nahm-fugacity-specialization}
\end{equation}
turns \eqref{eq:abelian-half-index} into the Nahm sum \eqref{eq:AD-Nahm-sum}.  The $q$-dependent part of this specialization is equivalently encoded by mixing the reference R-charge $R_0$ with the topological symmetries:
\begin{equation}
 R_H
 :=
 R_0-
 \sum_{a=1}^{4n-1}B^{(4n-2)}_aJ_a.
 \label{eq:Nahm-R-mixing}
\end{equation}
Here $R_0$ is chosen so that the elementary chiral multiplets have vanishing $R_0$ charge.  Equations \eqref{eq:3d-gauge-group}--\eqref{eq:Nahm-R-mixing} determine the gauge group, matter content, CS couplings, and the R-symmetry assignment read from the Nahm sum.
We next identify a monopole superpotential compatible with the Coulomb branch spectrum expected from the R-twisted reduction.

\subsection{The monopole superpotential}

\subsubsection{Constraints from the 4d Coulomb branch}
\label{subsec:monopole-constraints}

The Coulomb branch scaling dimensions of the $(A_2,D_{3n-2})$ theory are \cite{Xie:2012hs,Wang:2015mra}
\begin{equation}
 \left\{
 \frac{2k}{2n-1}\ \middle|\ k=n,n+1,\ldots,3n-3
 \right\}
 \sqcup
 \left\{
 \frac{2k}{2n-1}\ \middle|\ k=n,n+1,\ldots,2n-2
 \right\}
 \sqcup
 \left\{
 \frac{3n-2}{2n-1}
 \right\}.
 \label{eq:CB-spectrum-family}
\end{equation}
Here $\sqcup$ denotes a union with multiplicities retained.  There is precisely one operator of integer dimension, namely the dimension-two operator obtained by setting $k=2n-1$ in the first set.  The operators of fractional $U(1)_r$ charge are removed by the R-twisted compactification, whereas this dimension-two operator survives \cite{Hamachika:2026whv}.

This conclusion also has a direct semiclassical interpretation in the conformal gauging frame.  The surviving dimension-two operator is the quadratic Casimir of the 4d $\mathfrak{so}(3) \simeq \mathfrak{su}(2)$ vector multiplet scalar $\mathrm{Tr}\phi^2_{\rm4d}$.  At a generic point on the Coulomb branch, $SO(3)$ is broken to $U(1)$.  After the R-twisted circle compactification, the resulting 3d $\mathcal N=4$ abelian vector multiplet contains three real scalars, arising from the two real components of $\phi_{\rm 4d}$ and the circle component of the gauge field, together with the periodic dual photon.  These four real degrees of freedom give a Coulomb branch of quaternionic dimension one, or equivalently complex dimension two.  Quantum corrections may modify its geometry, but not this expected dimension.

This expectation constrains the monopole deformation of the ultraviolet $U(1)^{4n-1}$ CS matter theory.  The superpotential must be built from gauge invariant chiral monopole operators of $R_H=2$.  
Motivated by the expected complex two-dimensional Coulomb branch and by the \(n=2\) construction \cite{Hamachika:2026whv}, we impose that the superpotential leave two independent \(\mathcal N=2\) topological symmetries unbroken. In the infrared, one linear combination of these symmetries is expected to enter the R-symmetry, while an independent combination becomes a flavor symmetry acting on the Coulomb branch.

For later use, set
\begin{equation}
 N:=4n-1,
 \qquad
 L:=N-2=4n-3.
 \label{eq:N-L-definitions}
\end{equation}
We denote the standard basis of $\mathbb Z^N$ by $\bm e_a$, $a=1,\ldots,N$.  
We require the two topological $U(1)$ symmetries associated with the following charge vectors to be preserved:
\begin{align}
 \bm v^{(1)}
 &=
 \bigl(2n-2,\ 2n-1,\ 1,2,\ldots,L\bigr),
 \nonumber\\
 \bm v^{(2)}
 &=
 \bigl(2n-1,\ 2n-2,\ 1,2,\ldots,L\bigr).
 \label{eq:preserved-vectors-general}
\end{align}
They obey
\begin{equation}
 B^{(4n-2)}=\bm v^{(1)}+\bm v^{(2)}.
 \label{eq:B-v-sum}
\end{equation}
At $n=2$, equation \eqref{eq:preserved-vectors-general} reduces to the two vectors $(2,3,1,2,3,4,5)$ and $(3,2,1,2,3,4,5)$ appearing in \cite{Hamachika:2026whv}.

\subsubsection{Gauge invariant monopoles of $R_H=2$}
\label{subsec:RH2-monopoles}

Consider a dressed monopole operator
\begin{equation}
 \left(\prod_{a=1}^{N}\phi_a^{d_a}\right)V_{\bm m},
 \qquad
 \bm m\in\mathbb Z^N,
 \qquad
 \bm d\in\mathbb Z_{\geq0}^N,
 \label{eq:general-dressed-monopole}
\end{equation}
where $\phi_a$ is the scalar in the chiral multiplet charged under $U(1)_a$, and $V_{\boldsymbol{m}}$ is a bare monopole operator
with a magnetic charge $\boldsymbol{m}$.  In the effective CS convention used in the half-index, gauge invariance requires
\begin{equation}
 K^{(4n-2)}\bm m+\bm d=\bm p(\bm m),
 \qquad
 p_a(\bm m):=\max(m_a,0),
 \label{eq:gauge-invariance-general}
\end{equation}
with the BPS dressing condition $d_am_a=0$ for every $a$.  This is the direct generalization of the $n=2$ convention used 
in \cite{Hamachika:2026whv}.  The $R_H$-charge of the dressed monopole operator \eqref{eq:general-dressed-monopole} is
\begin{equation}
 R_H
 =
 \sum_{a=1}^{N}p_a(\bm m)
 -B^{(4n-2)}\!\cdot\bm m.
 \label{eq:RH-general-monopole}
\end{equation}
In particular, for monopoles neutral under both symmetries in \eqref{eq:preserved-vectors-general}, equation \eqref{eq:B-v-sum} gives
\begin{equation}
 R_H=\sum_{a=1}^{N}p_a(\bm m).
 \label{eq:RH-neutral-monopole}
\end{equation}

We now introduce $L=4n-3$ magnetic charge vectors.  The first is
\begin{equation}
 \bm m^{(1)}
 =
 \bm e_1+\bm e_2-\bm e_N.
 \label{eq:monopole-m1-general}
\end{equation}
The remaining $L-1=4n-4$ vectors form a discrete second-difference chain:
\begin{align}
 \bm m^{(2)}
 &=2\bm e_3-\bm e_4,
 \label{eq:monopole-m2-general}\\
 \bm m^{(r+1)}
 &=-\bm e_{r+1}+2\bm e_{r+2}-\bm e_{r+3},
 &&2\leq r\leq L-1.
 \label{eq:monopole-chain-general}
\end{align}

A direct multiplication by the CS matrix gives
\begin{equation}
 K^{(4n-2)}\bm m^{(k)}
 =
 \bm p\bigl(\bm m^{(k)}\bigr),
 \qquad
 k=1,\ldots,L.
 \label{eq:bare-monopole-gauge-invariance}
\end{equation}
To see this uniformly, write the lower-right block of $K^{(4n-2)}$ as
$H_{st}=2\min(s,t)$, $1\leq s,t\leq L$, and denote the standard basis of
the space $\mathbb Z^L$ by $\bm\epsilon_s$.  Under the natural
embedding of $\mathbb Z^L$ into the full magnetic-charge lattice,
$\bm\epsilon_s$ is identified with $\bm e_{s+2}$, $s=1,\ldots,L$.  Then
\begin{equation}
 H(2\bm\epsilon_1-\bm\epsilon_2)=2\bm\epsilon_1,
 \qquad
 H(-\bm\epsilon_{r-1}+2\bm\epsilon_r-\bm\epsilon_{r+1})
 =2\bm\epsilon_r,
 \quad 2\leq r\leq L-1.
 \label{eq:tail-second-difference-identity}
\end{equation}
The mixed entries in the first two rows of $K^{(4n-2)}$ vanish on the same combinations because their dependence on $s$ is linear.
 Hence all these operators are gauge invariant bare monopoles; no chiral dressing is required.  Explicitly,
\begin{equation}
 \bm p\bigl(\bm m^{(1)}\bigr)=\bm e_1+\bm e_2,
 \qquad
 \bm p\bigl(\bm m^{(r+1)}\bigr)=2\bm e_{r+2},
 \qquad
 1\leq r\leq L-1.
 \label{eq:positive-parts-candidates}
\end{equation}
They are neutral under both preserved topological symmetries,
\begin{equation}
 \bm v^{(i)}\!\cdot\bm m^{(k)}=0,
 \qquad
 i=1,2,
 \qquad
 k=1,\ldots,L,
 \label{eq:candidate-neutrality}
\end{equation}
where neutrality of the chain monopoles follows because the second difference
vectors annihilate the linear sequence $(1,2,\ldots,L)$.  Equations \eqref{eq:RH-neutral-monopole} and \eqref{eq:positive-parts-candidates} then imply
\begin{equation}
 R_H\bigl(V_{\bm m^{(k)}}\bigr)=2
 \qquad
 (k=1,\ldots,L).
 \label{eq:candidate-RH-two}
\end{equation}

\subsubsection{The monopole superpotential and its uniqueness}
\label{subsec:superpotential-uniqueness}

Let us consider a monopole superpotential
\begin{equation}
 W_n
 =
 \sum_{k=1}^{4n-3}V_{\bm m^{(k)}}
 \label{eq:general-monopole-superpotential}
\end{equation}
with the magnetic charges given in \eqref{eq:monopole-m1-general}--\eqref{eq:monopole-chain-general}.  Since these $4n-3$ vectors are linearly independent, their common orthogonal complement in the topological charge lattice has dimension two.  Equations \eqref{eq:candidate-neutrality} show that it is spanned by $\bm v^{(1)}$ and $\bm v^{(2)}$.  Consequently, for generic nonzero coefficients, the superpotential \eqref{eq:general-monopole-superpotential} breaks the ultraviolet topological symmetry as
\begin{equation}
 U(1)^{4n-1}_{J}
 \longrightarrow
 U(1)_{\bm v^{(1)}}\times U(1)_{\bm v^{(2)}}.
 \label{eq:topological-breaking-general}
\end{equation}
For $n=2$, the $4n-3=5$ terms in \eqref{eq:general-monopole-superpotential} reproduce precisely the monopole superpotential of \cite{Hamachika:2026whv}.

We can also establish the uniqueness of monopole superpotential  within the class of gauge invariant  monopole operators of $R_H=2$ that preserve both symmetries in \eqref{eq:preserved-vectors-general}.  Neutrality under these two symmetries is equivalent to
\begin{equation}
 m_1=m_2,
 \qquad
 Lm_1+\sum_{r=1}^{L}r\,m_{r+2}=0.
 \label{eq:neutrality-conditions-general}
\end{equation}
Every integer vector satisfying \eqref{eq:neutrality-conditions-general} has a unique expansion
\begin{equation}
 \bm m
 =
 c_0\bm m^{(1)}
 +\sum_{r=1}^{L-1}c_r\bm m^{(r+1)},
 \qquad
 c_r\in\mathbb Z.
 \label{eq:monopole-basis-expansion}
\end{equation}
The coefficients are determined explicitly by
\begin{equation}
 c_0=m_1,
 \qquad
 c_r
 =
 r m_1+
 \sum_{s=1}^{L}\min(r,s)m_{s+2},
 \qquad
 1\leq r\leq L-1,
 \label{eq:monopole-basis-coefficients}
\end{equation}
which also makes the integrality and uniqueness of the expansion manifest.  In this basis the CS quadratic form of the half-index becomes
\begin{align}
 \bm m^{\mathsf T}K^{(4n-2)}\bm m
 =2\Bigl[
 c_0^2+c_1^2
 +\sum_{r=1}^{L-2}(c_r-c_{r+1})^2
 +c_{L-1}^2
 \Bigr].
 \label{eq:quadratic-form-general-basis}
\end{align}
Multiplying the gauge-invariance condition \eqref{eq:gauge-invariance-general} by $\bm m^{\mathsf T}$ and using $d_am_a=0$ gives
\begin{equation}
 \bm m^{\mathsf T}K^{(4n-2)}\bm m
 =
 \sum_{a=1}^{N}p_a(\bm m)^2.
 \label{eq:quadratic-positive-relation}
\end{equation}
For a neutral operator with $R_H=2$, equation \eqref{eq:RH-neutral-monopole} says $\sum_a p_a=2$.  The right-hand side of \eqref{eq:quadratic-positive-relation} is therefore either $2$, when the positive part consists of two entries equal to one, or $4$, when it consists of one entry equal to two.

Define
\begin{equation}
 S(\bm c)
 :=
 c_1^2
 +\sum_{r=1}^{L-2}(c_r-c_{r+1})^2
 +c_{L-1}^2.
 \label{eq:chain-quadratic-form}
\end{equation}
For a nonzero integer sequence $(c_1,\ldots,c_{L-1})$, one has $S(\bm c)\geq2$.  Equality holds precisely when the sequence is equal to $+1$ or $-1$ on one nonempty consecutive interval and vanishes outside it.  If the quadratic form in \eqref{eq:quadratic-form-general-basis} equals two, this implies $c_0=1$ and $c_r=0$ for $r\geq1$, giving $\bm m^{(1)}$; the choice $c_0=-1$ has only one positive component and does not have $R_H=2$.  If the quadratic form equals four, then $c_0=0$ and $S(\bm c)=2$.  A positive interval of length one gives precisely one of the chain vectors \eqref{eq:monopole-m2-general}--\eqref{eq:monopole-chain-general}.  An interval of greater length, or a negative interval, has a positive part incompatible with \eqref{eq:quadratic-positive-relation} and $\sum_a p_a=2$.  Finally, for every surviving solution equation \eqref{eq:bare-monopole-gauge-invariance} forces $\bm d=0$.

It follows that the operators appearing in
\eqref{eq:general-monopole-superpotential} are the only gauge invariant
monopole operators of $R_H=2$ neutral under both
$U(1)_{\bm v^{(1)}}$ and $U(1)_{\bm v^{(2)}}$.
Thus the set of monopole operators in the superpotential is uniquely fixed by
the two preserved topological symmetries and the $R_H=2$ condition, while
their nonzero coefficients are not determined by this argument.

\subsection{Coulomb branch chiral ring}
\label{sec:Coulomb chiral-ring}

We next determine the chiral operators that can parameterize the Coulomb branch after the monopole deformation \eqref{eq:general-monopole-superpotential}.  In an infrared $\mathcal N=4$ description, Coulomb branch operators are neutral under $SU(2)_H$ and hence have $R_H=0$.  We therefore classify all gauge-invariant dressed monopole operators of $R_H=0$ in the ultraviolet CS matter theory.  The result is independent of $n$: the ring is generated by three bare monopoles obeying the $A_1$ surface-singularity relation.

\subsubsection{Classification of the $R_H=0$ monopoles}
\label{subsec:RH0-classification}

Consider again the dressed monopole operator \eqref{eq:general-dressed-monopole}, subject to the gauge-invariance and BPS dressing conditions
\begin{equation}
 K^{(4n-2)}\bm m+\bm d=\bm p(\bm m),
 \qquad
 d_am_a=0.
 \label{eq:RH0-gauge-invariance}
\end{equation}
A useful special property of the matrix \eqref{eq:Kp-components} is that its last row is precisely the vector \eqref{eq:Bp-vector}:
\begin{equation}
 \bigl(K^{(4n-2)}\bigr)_{Na}=B^{(4n-2)}_a,
 \qquad N=4n-1.
 \label{eq:last-row-is-B}
\end{equation}
Taking the last component of \eqref{eq:RH0-gauge-invariance} and substituting it into \eqref{eq:RH-general-monopole}, we obtain
\begin{equation}
 R_H
 =
 \sum_{a=1}^{N}p_a(\bm m)-B^{(4n-2)}\!\cdot\bm m
 =
 \sum_{a=1}^{N-1}p_a(\bm m)+d_N.
 \label{eq:RH-last-row-formula}
\end{equation}
Consequently, $R_H=0$ implies
\begin{equation}
 m_a\leq0\quad (a<N),
 \qquad
 d_N=0.
 \label{eq:RH0-sign-condition}
\end{equation}
The only solution with $m_N\leq0$ is the trivial operator.  Indeed, in that case all components of $\bm m$ are non-positive.  Multiplication of \eqref{eq:RH0-gauge-invariance} by $\bm m^{\mathsf T}$, together with $\bm m\cdot\bm d=0$, gives $\bm m^{\mathsf T}K^{(4n-2)}\bm m=0$.  To see the consequence, write $u=m_1+m_2$ and $x_r=m_{r+2}$, $r=1,\ldots,L$.  The quadratic form admits the factorization
\begin{equation}
 \bm m^{\mathsf T}K^{(4n-2)}\bm m
 =
 \frac12u^2
 +2\sum_{s=1}^{L}
 \left(\frac u2+\sum_{r=s}^{L}x_r\right)^2.
 \label{eq:K-positive-factorization}
\end{equation}
It is therefore positive semi-definite, with kernel spanned by $\bm e_1-\bm e_2$.  This kernel has no nonzero representative with all components non-positive.  Hence a nontrivial $R_H=0$ monopole operator must have $m_N >0$. We denote this positive integer by
\begin{equation}
 m_N=\nu>0.
 \label{eq:nu-positive}
\end{equation}

To solve the remaining conditions, write
\begin{equation}
 u:=m_1+m_2,
 \qquad
 x_r:=m_{r+2},
 \qquad r=1,\ldots,L,
 \qquad L=4n-3,
 \label{eq:u-x-definitions}
\end{equation}
and define
\begin{equation}
 A:=\bigl(K^{(4n-2)}\bm m\bigr)_1
   =\bigl(K^{(4n-2)}\bm m\bigr)_2,
 \qquad
 y_s:=\bigl(K^{(4n-2)}\bm m\bigr)_{s+2},
 \qquad y_0:=0.
 \label{eq:A-y-definitions}
\end{equation}
The explicit matrix gives
\begin{equation}
 B^{(4n-2)}\!\cdot\bm m=2A-u.
 \label{eq:B-2A-u}
\end{equation}
The first two components of \eqref{eq:RH0-gauge-invariance} give $A+d_1=A+d_2=0$, so $A\leq0$.  Combining the last component with \eqref{eq:B-2A-u} gives $\nu=2A-u$.  If $u=0$, then $m_1=m_2=0$ and hence $\nu=2A\leq0$, contradicting \eqref{eq:nu-positive}.  Thus $u<0$.  At least one of $m_1,m_2$ is negative, and the BPS dressing condition forces the corresponding dressing exponent to vanish; its gauge-invariance equation then gives $A=0$.  The last component of \eqref{eq:RH0-gauge-invariance} therefore yields
\begin{equation}
 u=-\nu.
 \label{eq:u-minus-nu}
\end{equation}
For the remaining components, gauge invariance and the BPS condition imply
\begin{equation}
 y_s\leq0,
 \qquad
 x_s\leq0,
 \qquad
 x_sy_s=0,
 \qquad 1\leq s<L,
 \label{eq:complementarity-tail}
\end{equation}
while $x_L=y_L=\nu$.  Moreover, direct subtraction of adjacent rows of the CS matrix yields
\begin{equation}
 y_{s+1}-2y_s+y_{s-1}=-2x_s\geq0,
 \qquad 1\leq s<L.
 \label{eq:discrete-convexity}
\end{equation}
Thus the first differences $\delta_s:=y_s-y_{s-1}$ are nondecreasing.  Suppose that $y_j<0$ for some $j<L$, and choose the smallest such $j$.  If $j>1$, then $y_{j-2}=y_{j-1}=0$, so $\delta_{j-1}=0$ while $\delta_j<0$, contradicting monotonicity.  If $j=1$, complementarity gives $x_1=0$ and hence $\delta_2=\delta_1<0$.  Repeating this argument along the maximal negative interval makes the sequence affine with negative slope, so it cannot reach the positive endpoint $y_L=\nu$.  Therefore
\begin{equation}
 y_s=0,
 \qquad 1\leq s<L.
 \label{eq:y-vanishing}
\end{equation}
Equation \eqref{eq:discrete-convexity} now determines the magnetic charge uniquely up to the split between its first two components:
\begin{equation}
 x_1=\cdots=x_{L-2}=0,
 \qquad
 x_{L-1}=-\frac \nu2,
 \qquad
 x_L=\nu.
 \label{eq:x-solution}
\end{equation}
Integrality requires $\nu=2k$ with $k\in\mathbb Z_{>0}$.  All dressing exponents vanish, and the complete set of nontrivial gauge invariant dressed monopoles of $R_H=0$ is therefore
\begin{equation}
 \bm m(k,a)
 =
 -a\bm e_1-(2k-a)\bm e_2-k\bm e_{N-1}+2k\bm e_N,
 \qquad
 k\geq1,
 \quad
 0\leq a\leq2k.
 \label{eq:all-RH0-monopoles}
\end{equation}
In particular, the classification excludes additional scalar-dressed generators.

\subsubsection{Generators and chiral ring relation of the Coulomb branch}
\label{subsec:A1-ring}

The semigroup \eqref{eq:all-RH0-monopoles} is generated by the three magnetic charges
\begin{align}
 \bm m_X
 &=
 -2\bm e_2-\bm e_{N-1}+2\bm e_N,
 \nonumber\\
 \bm m_Y
 &=
 -2\bm e_1-\bm e_{N-1}+2\bm e_N,
 \nonumber\\
 \bm m_Z
 &=
 -\bm e_1-\bm e_2-\bm e_{N-1}+2\bm e_N.
 \label{eq:XYZ-magnetic-charges}
\end{align}
We denote the corresponding bare monopoles by
\begin{equation}
 X:=V_{\bm m_X},
 \qquad
 Y:=V_{\bm m_Y},
 \qquad
 Z:=V_{\bm m_Z}.
 \label{eq:XYZ-definitions}
\end{equation}
Indeed, if the integer $a$ in \eqref{eq:all-RH0-monopoles} is even, then
\begin{equation}
 \bm m(k,a)
 =
 \left(k-\frac a2\right)\bm m_X
 +\frac a2\bm m_Y,
 \label{eq:even-a-decomposition}
\end{equation}
whereas for odd $a$,
\begin{equation}
 \bm m(k,a)
 =
 \frac{2k-a-1}{2}\bm m_X
 +\frac{a-1}{2}\bm m_Y
 +\bm m_Z.
 \label{eq:odd-a-decomposition}
\end{equation}
We next determine the chiral ring relation among $X$, $Y$, and $Z$.
Their magnetic charges satisfy
\begin{equation}
 \bm m_X+\bm m_Y=2\bm m_Z.
 \label{eq:magnetic-XY-Z2}
\end{equation}
A possible relation therefore takes the form
\begin{equation}
 XY=f(\phi)Z^2,
\end{equation}
where $f(\phi)$ is a polynomial in the elementary chiral multiplet scalars.

We now show that no scalar dressing is required. Since $\bm m_X$ and
$\bm m_Y$ have no components of opposite sign,
\begin{equation}
 \bm p(\bm m_X)+\bm p(\bm m_Y)
 =
 \bm p(\bm m_X+\bm m_Y).
 \label{eq:XY-positive-part-additivity}
\end{equation}
Using $K^{(4n-2)}\bm m=\bm p(\bm m)$ for the three bare monopole generators,
we obtain
\begin{equation}
 K^{(4n-2)}(\bm m_X+\bm m_Y)
 =
 \bm p(\bm m_X+\bm m_Y).
 \label{eq:XY-product-bare-gauge-invariance}
\end{equation}
Thus the monopole operator of charge $\bm m_X+\bm m_Y$ is gauge invariant
without scalar dressing. The same argument applies to the monopole operator
of charge $2\bm m_Z$.

Moreover, there is no nonconstant gauge invariant polynomial constructed
solely from the elementary chiral fields, since each $\phi_a$ is charged
under its own $U(1)_a$ gauge factor. Hence $f(\phi)$ is a constant.
Provided that the corresponding monopole OPE coefficient is nonzero, this
constant can be absorbed into the normalization of the generators, giving
the candidate chiral ring relation
\begin{equation}
 XY=Z^2.
 \label{eq:XY-Z2-general}
\end{equation}


We consequently obtain the candidate Coulomb branch chiral ring
\begin{equation}
 \mathbb C[\mathcal M_C]
 \simeq
 \frac{\mathbb C[X,Y,Z]}{(XY-Z^2)},
 \qquad
 \mathcal M_C\simeq\mathbb C^2/\mathbb Z_2.
 \label{eq:CB-ring-general}
\end{equation}
This result is independent of $n$.  It is consistent with the fact that every member of the 4d family has precisely one Coulomb branch generator with integer scaling dimension, which produces a quaternionic one-dimensional Coulomb branch after the R-twisted reduction.

The charges under the two topological symmetries preserved by the superpotential are
\begin{equation}
 \begin{array}{c|ccc}
  &X&Y&Z\\ \hline
 U(1)_{\bm v^{(1)}}&0&2&1\\
 U(1)_{\bm v^{(2)}}&2&0&1
 \end{array}.
 \label{eq:XYZ-v-charges}
\end{equation}
It is useful to introduce the average and difference
\begin{equation}
 J_{\mathrm{av}}:=\frac12\bigl(J_{\bm v^{(1)}}+J_{\bm v^{(2)}}\bigr),
 \qquad
 J_F:=J_{\bm v^{(2)}}-J_{\bm v^{(1)}}.
 \label{eq:Jav-JF-definitions}
\end{equation}
Then $X,Y,Z$ all have $J_{\mathrm{av}}=1$, while their $J_F$ charges are $+2,-2,0$, respectively.  The generators therefore form the weight triplet of an $SU(2)$ symmetry, and \eqref{eq:XY-Z2-general} is its invariant quadratic relation.  Equivalently, the refined Hilbert series is
\begin{equation}
 H_{\mathcal M_C}(t,x)
 =
 \frac{1-t^2}
 {(1-tx^2)(1-t)(1-tx^{-2})}.
 \label{eq:CB-Hilbert-series}
\end{equation}
It admits the character expansion
\begin{equation}
 H_{\mathcal M_C}(t,x)
 =
 \sum_{\ell=0}^{\infty}
 \chi_{\mathbf{2\ell+1}}(x)t^{\ell},
 \label{eq:CB-Hilbert-character-expansion}
\end{equation}
where $\chi_{\mathbf{2\ell+1}}(x)$ is the character of the spin-$\ell$ representation of $SU(2)$.  This makes manifest the expected enhancement of the surviving Coulomb branch flavor $U(1)$ to $SU(2)$.

\subsubsection{Compatibility with the monopole superpotential}
\label{subsec:CB-superpotential-compatibility}

It remains to check semi-classically that the directions generated by $X,Y,Z$ are compatible with the F-term constraints of \eqref{eq:general-monopole-superpotential}.  Write a monopole operator as
\begin{equation}
 V_{\bm m}\sim\exp(\bm m\cdot\bm\varphi),
 \label{eq:semiclassical-monopole-exponential}
\end{equation}
where $\bm\varphi$ is the complex combination of  the vector multiplet scalars and dual photons.  Allowing generic nonzero coefficients $\lambda_k$, the superpotential and its F-term equation take the form
\begin{align}
 W_n
 &\sim
 \sum_{k=1}^{N-2}\lambda_k
 \exp\bigl(\bm m^{(k)}\cdot\bm\varphi\bigr),
 \nonumber\\
 \bm 0
 &=
 \frac{\partial W_n}{\partial\bm\varphi}
 =
 \sum_{k=1}^{N-2}\lambda_k\bm m^{(k)}
 \exp\bigl(\bm m^{(k)}\cdot\bm\varphi\bigr).
 \label{eq:semiclassical-monopole-F-term}
\end{align}
Because the $N-2$ magnetic charges $\bm m^{(k)}$ are linearly independent, \eqref{eq:semiclassical-monopole-F-term} requires each exponential coefficient to vanish.  This cannot occur at finite $\bm\varphi$; rather, the supersymmetric Coulomb branch locus is reached asymptotically through
\begin{equation}
 \operatorname{Re}\bigl(\bm m^{(k)}\cdot\bm\varphi\bigr)
 \longrightarrow-\infty,
 \qquad
 k=1,\ldots,N-2.
 \label{eq:monopole-F-term-asymptotic}
\end{equation}
Thus these $N-2$ logarithmic combinations cannot remain finite on the semiclassical Coulomb branch, as in the $n=2$ construction of \cite{Hamachika:2026whv}.

On the other hand, the charge matrix of $\bm m_X$ and $\bm m_Y$ under the two preserved topological symmetries is
\begin{equation}
 \begin{pmatrix}
  \bm v^{(1)}\!\cdot\bm m_X & \bm v^{(1)}\!\cdot\bm m_Y\\
  \bm v^{(2)}\!\cdot\bm m_X & \bm v^{(2)}\!\cdot\bm m_Y
 \end{pmatrix}
 =
 \begin{pmatrix}
  0&2\\
  2&0
 \end{pmatrix},
 \label{eq:XY-preserved-charge-matrix}
\end{equation}
which is non-degenerate.  Since the $\bm m^{(k)}$ span the common kernel of $\bm v^{(1)}$ and $\bm v^{(2)}$, it follows that
\begin{equation}
 \bigl\{\bm m^{(1)},\ldots,\bm m^{(N-2)},\bm m_X,\bm m_Y\bigr\}
 \label{eq:full-magnetic-basis}
\end{equation}
forms a basis of $\mathbb R^N$.  The two combinations that can remain finite while the limits \eqref{eq:monopole-F-term-asymptotic} are taken may be chosen as 
$\bm m_X\cdot\bm\varphi$ and $\bm m_Y\cdot\bm\varphi $
with $X\sim e^{\bm m_X\cdot\bm\varphi}$, $Y\sim e^{\bm m_Y\cdot\bm\varphi}$, and, because $2\bm m_Z=\bm m_X+\bm m_Y$, $Z\sim e^{(\bm m_X+\bm m_Y)\cdot\bm\varphi/2}$.  This does not by itself constitute an independent quantum derivation of the full moduli space.  It provides a semiclassical consistency check that the monopole superpotential lifts $N-2$ logarithmic directions while remaining compatible with the two complex directions and the chiral ring \eqref{eq:CB-ring-general} found above.

At $n=2$, equations
\eqref{eq:XYZ-magnetic-charges}--\eqref{eq:CB-ring-general}
reduce to the chiral ring analysis of \cite{Hamachika:2026whv}.
The general $n$ classification therefore extends the
$\mathbb C^2/\mathbb Z_2$ Coulomb branch found for the R-twisted reduction
of $(A_2,D_4)$ to the full $(A_2,D_{3n-2})$ family.

\subsection{Superconformal index and supersymmetry enhancement}
\label{sec:3d-superconformal-index}

The Coulomb branch analysis determines a natural candidate for the infrared
superconformal R-symmetry.  Recall the average topological charge introduced in
\eqref{eq:Jav-JF-definitions},
\begin{equation}
 J_{\mathrm{av}}
 =\frac12\left(J_{\bm v^{(1)}}+J_{\bm v^{(2)}}\right)
 =\frac12\sum_{a=1}^{N}B^{(4n-2)}_aJ_a.
 \label{eq:Jav-B-relation}
\end{equation}
The moment map candidates $X,Y,Z$ have $R_H=0$ and
$J_{\mathrm{av}}=1$.  They therefore have superconformal charge one if we
identify
\begin{equation}
 \frac{R_C-R_H}{2}=J_{\mathrm{av}},
 \qquad
 R_{\mathrm{SC}}
 =\frac{R_H+R_C}{2}
 =R_H+J_{\mathrm{av}}
 =R_0-\frac12\sum_{a=1}^{N}B^{(4n-2)}_aJ_a.
 \label{eq:RSC-general}
\end{equation}
Here $R_H$ (resp. $R_C$) is a Cartan generator of the infrared $SU(2)_H$ (resp. $SU(2)_C$) R-symmetry acting on Higgs (resp. Coulomb) branch.
$R_{\rm SC}$ is a superconformal $\mathcal{N}=2$ R-symmetry generator.
 
This is the direct generalization of the charge assignment used for the
$(A_2,D_4)$ theory in \cite{Hamachika:2026whv}.  It also assigns
$R_{\mathrm{SC}}=2$ to every term in the monopole superpotential
\eqref{eq:general-monopole-superpotential}.

We first evaluate the superconformal index for $n=3$.
We use the superconformal index
\begin{equation}
 \mathcal I_{\mathrm{SCI}}(\mathfrak q,T,x)
 =\Tr_{\mathcal{H}_{S^2}}(-1)^F
 \mathfrak q^{j+R_{\mathrm{SC}}/2}
 T^{J_{\mathrm{av}}}x^{-J_F},
 \label{eq:SCI-definition}
\end{equation}
where the sign in the last exponent is chosen so that a monopole operator of magnetic
charge $\bm m$ is weighted by $x^{m_2-m_1}$.  Reversing this convention is the
Weyl reflection $x\leftrightarrow x^{-1}$ and does not affect the result.  
For the CS matter theory defined in
\eqref{eq:3d-gauge-group}, the  localization formula 
\cite{Kim:2009wb,Imamura:2011su} gives
\begin{align}
 \mathcal I_{\mathrm{SCI}}(\mathfrak q,T,x)
 & =
 \sum_{\bm m\in\mathbb Z^{4n-1}}
 \oint\prod_{a=1}^{4n-1}\frac{dz_a}{2\pi iz_a}
 \prod_{a=1}^{4n-1}z_a^{(K^{(4n-2)}\bm m)_a}
 \left(\mathfrak q^{-1/2}T\right)^{\frac12B^{(4n-2)}\cdot\bm m}
 x^{m_2-m_1}
 \nonumber\\
 &\quad\times
 \prod_{a=1}^{4n-1}
 \left(-\mathfrak q^{1/2}z_a^{-1}\right)^{p_a(\bm m)}
 \frac{
  \left(z_a^{-1}\mathfrak q^{1+|m_a|/2};\mathfrak q\right)_\infty
 }{
  \left(z_a\mathfrak q^{|m_a|/2};\mathfrak q\right)_\infty
 },
 \label{eq:SCI-localization-general}
\end{align}
where $p_a(\bm m)=\max(m_a,0)$ and
\begin{equation}
 (y;\mathfrak q)_\infty:=\prod_{r=0}^{\infty}(1-y\mathfrak q^r).
\end{equation}

For $n=3$, so that $N=11$ and
$B^{(10)}=(9,9,2,4,\ldots,18)$, we obtain
\begin{align}
 \mathcal I_{\mathrm{SCI}}(\mathfrak q,T,x)
 & =1
 +T\chi_{\mathbf3}(x)\mathfrak q^{1/2}
 +\left[-1-\chi_{\mathbf3}(x)
       +T^2\chi_{\mathbf5}(x)\right]\mathfrak q
 \nonumber\\
 &\quad
 +\left[
 T+T^{-1}-T\chi_{\mathbf5}(x)
 +T^3\chi_{\mathbf7}(x)
 \right]\mathfrak q^{3/2}
 +O(\mathfrak q^2).
 \label{eq:n3-SCI-expansion}
\end{align}
Note that \eqref{eq:n3-SCI-expansion} agrees, through this order, with the superconformal index of the $n=2$ theory
found in \cite{Hamachika:2026whv}. In the next subsection, we will show that this low-order expansion is in fact universal for all $n\geq2$.

Several terms have direct multiplet interpretations.  The contribution
$T\chi_{\mathbf3}(x)\mathfrak q^{1/2}$ comes from the three operators
$X,Y,Z$, confirming that they have the quantum numbers of the
$SU(2)_C^{\mathrm{flavor}}$ moment maps.  At order $\mathfrak q$, the singlet
$-1$ is the current multiplet for $J_{\mathrm{av}}$, which is contained in the
$\mathcal N=4$ stress-tensor multiplet, while
$-\chi_{\mathbf3}(x)$ is the current multiplet of the enhanced
$SU(2)_C^{\mathrm{flavor}}$.  Most importantly, the terms
\begin{equation}
 (T+T^{-1})\mathfrak q^{3/2}
 \label{eq:extra-supercurrent-terms}
\end{equation}
are precisely the contributions expected from the extra supercurrent
multiplets that enhance the manifest $\mathcal N=2$ supersymmetry to
$\mathcal N=4$.  As usual, using $(-1)^{R_{\mathrm{SC}}}$ rather than
$(-1)^F$ changes the sign of these terms by the replacement
$\mathfrak q^{1/2}\mapsto-\mathfrak q^{1/2}$.

The following two limits give  further checks.  In the Higgs branch limit,
\begin{equation}
 \mathfrak q\longrightarrow0,
 \qquad
 T\mathfrak q^{-1/2}\ \text{fixed},
\end{equation}
no nonconstant contribution appears through the computed order.  This is
consistent with the expected exact result $\mathcal I_H=1$ and hence with a
trivial Higgs branch.  In the Coulomb branch limit,
\begin{equation}
 \mathfrak q\longrightarrow0,
 \qquad
 \mathfrak t:=T\mathfrak q^{1/2}\ \text{fixed},
\end{equation}
we find
\begin{equation}
 \mathcal I_C(\mathfrak t,x)
 =1+\chi_{\mathbf3}(x)\mathfrak t
  +\chi_{\mathbf5}(x)\mathfrak t^2
  +\chi_{\mathbf7}(x)\mathfrak t^3
  +O(\mathfrak t^4).
 \label{eq:n3-Coulomb limit}
\end{equation}
This agrees with the expansion of the refined Hilbert series
\eqref{eq:CB-Hilbert-series} of $\mathbb C^2/\mathbb Z_2$.

The index calculation therefore simultaneously supports the proposed
superconformal R-charge, the enhancement of the Coulomb branch flavor symmetry
to $SU(2)$, the triviality of the Higgs branch, and the enhancement to
$\mathcal N=4$ supersymmetry at the infrared fixed point.  It remains a
finite-order consistency test rather than a proof of the RG flow, but it is a
nontrivial check of the IR structure of the CS matter theory.

\subsubsection{Universal low order expression of the superconformal index}
\label{subsec:general-n-SCI}

We show that the superconformal index has the universal expansion
\eqref{eq:n3-SCI-expansion} through order $\mathfrak q^{3/2}$ for every
$n\geq2$. To this end, we first classify the magnetic sectors that can
contribute up to this order.

A direct classification in terms of the components of the magnetic charge
$\bm m\in\mathbb Z^N$ is cumbersome. It is more convenient to expand
$\bm m$ in terms of the monopole charges introduced above, for which the
action of the CS matrix takes a particularly simple form.

Set $ M=N-3=4n-4$. Using the monopole charges $\bm m^{(a)}$ introduced above together with the monopole charges $\bm m_X$ and 
$\bm m_Y$ in \eqref{eq:XYZ-magnetic-charges}, the action of 
the CS matrix takes the sparse form
\begin{align}
 K^{(4n-2)}\bm m^{(1)}&=\bm e_1+\bm e_2,
 &
 K^{(4n-2)}\bm m^{(r+1)}&=2\bm e_{r+2},
 \quad r=1,\ldots,M,
 \nonumber\\
 K^{(4n-2)}\bm m_X&=2\bm e_N,
 &
 K^{(4n-2)}\bm m_Y&=2\bm e_N.
 \label{eq:SCI-diagonal-magnetic-basis}
\end{align}
Every magnetic charge has a unique  expansion
\begin{equation}
 \bm m
 =a\bm m^{(1)}+\sum_{r=1}^{M}c_r\bm m^{(r+1)}
 +\frac{u}{2}\bm m_X+\frac{v}{2}\bm m_Y.
 \label{eq:general-sector-coordinates}
\end{equation}
For an integral magnetic charge, $a,u,v$ are integers, while the $c_r$ are
integers or half integral and are subject to additional integrality conditions.
The precise integrality conditions and
the finite classification used below are derived in
Appendix~\ref{app:low-order-sector-classification}.  Introduce
\begin{equation}
 s:=u+v=B^{(4n-2)}\cdot\bm m,
 \qquad
 c_0:=0,
 \qquad
 c_{M+1}:=\frac{s}{2}.
\end{equation}
The components of \eqref{eq:general-sector-coordinates} are then
\begin{align}
 m_1&=a-v,
 &m_2&=a-u,
 \nonumber\\
 m_{r+2}&=-c_{r-1}+2c_r-c_{r+1},
 &&r=1,\ldots,M,
 \nonumber\\
 m_N&=s-a-c_M.
 \label{eq:general-sector-components}
\end{align}
The flavor powers of the corresponding index sector are
\begin{equation}
 T^{s/2}x^{v-u}.
 \label{eq:sector-flavor-weights}
\end{equation}
With this parametrization, the problem is reduced to determining which values of \((a,c_r,u,v)\) can contribute through order \(\mathfrak q^{3/2}\).

We now determine which magnetic sectors can contribute through order $\mathfrak{q}^{3/2}$. For a fixed magnetic charge $\bm m$, consider the $z_b$ integral in the localization formula for the 
superconformal index. The one-loop determinant of the $b$-th chiral multiplet can be expanded as
\begin{align}
    \frac{(z_b^{-1} \mathfrak{q}^{1+|m_b|/2}; \mathfrak{q})_{\infty}}
    {(z_b\mathfrak{q}^{|m_b|/2};\mathfrak{q})_{\infty}}
    =\sum_{\rho, \sigma \ge 0}
    \frac{(-1)^{\sigma} z^{\rho-\sigma}_{b} 
    \mathfrak{q}^{\frac{1}{2}\rho |m_b|+\sigma(1+|m_b|/2)+\sigma(\sigma-1)/2}}{(\mathfrak{q}; \mathfrak{q})_{\rho}(\mathfrak{q}; \mathfrak{q})_{\sigma}}
\end{align}
Including the remaining \(z_b\)-dependent factors in the localization integrand, the total power of \(z_b\) is
\begin{align}
 (K^{(4n-2)}\bm m)_b-\max(m_b,0)+\rho-\sigma.
\end{align}
Since the contour integral over \(z_b\) extracts the constant term, a contribution survives only when
\begin{align}
 \rho-\sigma = \max(m_b,0)-(K^{(4n-2)}\bm m)_b.     
\end{align}
The lowest power of \(\mathfrak q\) that can appear in a fixed magnetic sector is obtained by minimizing the power of \(\mathfrak q\) in the one-loop determinants subject to the constant-term conditions above. We denote the corresponding degree in powers of \(\mathfrak q^{1/2}\) by \(D(\bm m)\), so that the contribution of the magnetic sector \(\bm m\) can start at order $\mathfrak q^{D(\bm m)/2}. $
Explicitly,
\begin{align}
    D(\bm m) =-\frac{s}{2}+\sum_{b=1}^N 
 \min_{\substack{\rho,\sigma\geq0\\
 \rho-\sigma={\rm max}(m_b,0)-(K^{(4n-2)} {\bm m})_b}}
 \left[
 {\rm max}(m_b,0)+(\rho+\sigma)|m_b|+2\sigma+\sigma(\sigma-1)
 \right].
 \label{eq:single-chiral-leading-degree}
\end{align}
Hence any magnetic sector that contributes through order
$\mathfrak q^{3/2}$ must satisfy
\begin{equation}
 D(\bm m)\leq3.
\end{equation}
We therefore classify all magnetic sectors satisfying this bound.
For \(n\geq3\), solving this inequality together with the integrality conditions described above gives the complete list below. Here  $\bm c=(c_1,\ldots,c_M)$,
\(\bm e_j^{(M)}\) is the \(j\)-th standard vector of \(\mathbb Z^M\), \(\bm 1_M=(1,\ldots,1)\), and data not displayed are zero:
\begin{equation}
 \begin{array}{c|l|c} 
 s & \text{allowed coordinates }(u,v;a;\bm c)&\text{number}\\ \hline -2&(-1,-1;-1;-\bm1_M)&1\\[1mm] 0&(-1,1;0;\bm0),\ (1,-1;0;\bm0), \ (0,0;a;\bm c)\text{ with }(a,\bm c) =(0,\bm0),(1,\bm0),(0,\bm e_j^{(M)})&M+4\\[1mm] 2&(-1,3;0;\bm0),\ (3,-1;0;\bm0), \ (u,2-u;0;\bm c),\ u=0,1,2,\ \bm c\in\{\bm0,\bm e_1^{(M)},\ldots, \bm e_{M-1}^{(M)}\}&3M+2\\[1mm] 4&(u,4-u;0;\bm0),\quad u=0,\ldots,4&5\\[1mm] 6&(u,6-u;0;\bm0),\quad u=0,\ldots,6&7
 \end{array}    
\end{equation}

For \(n\geq3\), all solutions satisfying \(D(\bm m)\leq3\) have integer \(c_r\), and the total number of contributing magnetic sectors is
\begin{align}
    1+(M+4)+(3M+2)+5+7=16n+3, \qquad n\geq3. 
\end{align} 
This gives a finite classification of all magnetic sectors that can
contribute through order $\mathfrak q^{3/2}$ without imposing a cutoff on
the components of $\bm m$.

For $n=2$, there are eight additional sectors in which some of the $c_r$
are half integral. These sectors have $D(\bm m)=5/2$ and occur at $s=1$;
they are listed explicitly in Appendix~\ref{app:n2-exceptional-sectors}.
Their leading contributions cancel separately at flavor weights
$T^{1/2}x$ and $T^{1/2}x^{-1}$, and their higher order contributions start
beyond $\mathfrak q^{3/2}$. Consequently, they do not modify the index
through the order considered here.

Expanding the one-loop determinants in the magnetic sectors listed above, and including the cancelling additional sectors for \(n=2\), gives
\begin{align} s=-2:&\qquad T^{-1}\mathfrak q^{3/2},\nonumber \\ s=0:&\qquad 1-\bigl[1+\chi_{\mathbf3}(x)\bigr]\mathfrak q, \nonumber\\ s=2:&\qquad T\chi_{\mathbf3}(x)\mathfrak q^{1/2} +T\bigl[1-\chi_{\mathbf5}(x)\bigr]\mathfrak q^{3/2}, \nonumber\\ s=4:&\qquad T^2\chi_{\mathbf5}(x)\mathfrak q,\nonumber\\ s=6:&\qquad T^3\chi_{\mathbf7}(x)\mathfrak q^{3/2}.
\end{align}
We therefore obtain the universal formula \(\eqref{eq:n3-SCI-expansion}\) through order \(\mathfrak q^{3/2}\) for every \(n\geq2\).

This universality is only a low order statement. The full superconformal indices retain their dependence on \(n\). For example, the \(n=2\) and \(n=3\) indices first differ at order \(\mathfrak q^{9/4}\):
\begin{align}
    \mathcal I_{\mathrm{SCI}}^{(n=2)} - \mathcal I_{\mathrm{SCI}}^{(n=3)} = -T^{-3/2}\chi_{\mathbf2}(x)\mathfrak q^{9/4} +O(\mathfrak q^{5/2}),
\end{align} 
Thus the universal low-order terms, including the extra-supercurrent contribution, do not imply that all members of the family flow to the same infrared theory.

\section{Conjectural 4d/3d charge map and Macdonald refinement}
\label{sec:Macdonald-refinement}

In the previous sections, we identified two topological symmetries preserved by
the monopole superpotential.  Their infrared interpretation is summarized by
\eqref{eq:Jav-JF-definitions} and \eqref{eq:RSC-general}, with $J_F$ acting as
the Cartan generator of the $SU(2)$ flavor symmetry on the Coulomb branch.
We now ask how the 4d $R$-symmetry charges are mapped to these
3d charges under the $R$-twisted reduction.

For the \(n=2\) theory, this question can be tested using the independently known four-dimensional Macdonald index. In \cite{Hamachika:2026whv}, a refinement of the 3d half-index was shown to reproduce this Macdonald index.  We use this
result as the main input for a conjectural extension to the full
$(A_2,D_{3n-2})$ family.  The Macdonald refinement constructed below should
therefore be regarded as a quantitative consequence of the proposed 4d/3d
charge map, rather than as an independent derivation of it.

\subsection{Three-dimensional charge gradings}
\label{subsec:3d-charge-gradings}

Let $\bm v^{(1)}$ and $\bm v^{(2)}$ be the two topological charge vectors
defined in \eqref{eq:preserved-vectors-general}.  The corresponding physical
topological charges are $J_{\bm v^{(1)}}$ and $J_{\bm v^{(2)}}$.

The relation between the physical topological charges and the gradings
appearing in the Nahm sum follows directly from the half-index convention.
The half-index \eqref{eq:abelian-half-index} contains
$\prod_a x_a^{-\ell_a}$, where $x_a$ is the fugacity for $J_a$.
Thus the half-index sector labelled by $\bm\ell$ carries topological charge
$J_a=-\ell_a$ in this convention, and hence
\begin{equation}
 \bm v^{(i)}\cdot\bm\ell
 =
 -J_{\bm v^{(i)}},
 \qquad i=1,2.
 \label{eq:Nahm-grading-topological-charge}
\end{equation}

Recall from \eqref{eq:Jav-JF-definitions} and \eqref{eq:RSC-general} that
\begin{equation}
 J_{\mathrm{av}}
 =
 \frac12\left(
 J_{\bm v^{(1)}}+J_{\bm v^{(2)}}
 \right)
 =
 \frac{R_C-R_H}{2},
 \qquad
 J_F
 =
 J_{\bm v^{(2)}}-J_{\bm v^{(1)}}.
 \label{eq:Macdonald-Jav-JF}
\end{equation}
Here $J_F$ is the Cartan generator of the $SU(2)$ flavor symmetry acting on
the Coulomb branch.  This identification is supported by the
superconformal index, its Higgs and Coulomb branch limits, and the
$\mathbb C^2/\mathbb Z_2$ Coulomb branch chiral ring found above.

It follows that
\begin{align}
 J_{\bm v^{(1)}}
 &=
 \frac{R_C-R_H-J_F}{2},
 &
 J_{\bm v^{(2)}}
 &=
 \frac{R_C-R_H+J_F}{2}.
 \label{eq:physical-topological-charge-identification}
\end{align}
These relations are statements entirely about the three-dimensional theory.
In particular, no four-dimensional Macdonald index is used in obtaining them.

\subsection{Conjectural 4d/3d charge identification}
\label{subsec:4d-3d-charge-map}

The four-dimensional Macdonald index is
\begin{equation}
 \mathcal I_{\mathrm{Mac}}(q,t)
 =
 \Tr
 (-1)^F
 q^{\Delta-2R-r}
 t^{R+r}\,,
 \label{eq:Macdonald-index-definition}
\end{equation}
where $\Delta$ is the scaling dimension and $R$ and $r$ are the Cartan
generators of $SU(2)_R$ and $U(1)_r$, respectively.
Writing
\begin{equation}
 t=qT,
\end{equation}
this becomes
\begin{equation}
 \mathcal I_{\mathrm{Mac}}(q,T)
 =
 \Tr
 (-1)^F
 q^{\Delta-R}
 T^{R+r}.
 \label{eq:Macdonald-index-T-convention}
\end{equation}
Thus the additional fugacity $T$ keeps track of the 4d charge
$R+r$.

For the $n=2$ theory, the refined three-dimensional half-index was compared
with the independently known Macdonald index of the $(A_2,D_4)$ theory in
\cite{Hamachika:2026whv}.  In the conventions used here, this matching
identifies the 4d grading $R+r$ with one of the two gradings
$-J_{\bm v^{(1)}}$ and $-J_{\bm v^{(2)}}$ appearing in the Nahm sum.

The two choices are related by an exact symmetry.  The exchange
$\ell_1\leftrightarrow \ell_2$ leaves the quadratic form and the linear term of the
Nahm sum invariant and exchanges
\begin{equation}
 \bm v^{(1)}
 \longleftrightarrow
 \bm v^{(2)}.
\end{equation}
On the 3d Coulomb branch this is the Weyl reflection of the
enhanced $SU(2)$ flavor symmetry, which exchanges
$J_{\bm v^{(1)}}$ and $J_{\bm v^{(2)}}$.

The two preserved topological symmetries $J_{\bm v^{(1)}}$ and
$J_{\bm v^{(2)}}$, together with the exchange symmetry
$\bm v^{(1)}\leftrightarrow\bm v^{(2)}$, are present for every $n\geq2$.
Motivated by the $n=2$ matching, we therefore conjecture the following charge
map for the $R$-twisted reduction of the full family:
\begin{equation}
 R+r
 \quad\longleftrightarrow\quad
 -J_{\bm v^{(1)}}
 =
 \frac{R_H-R_C+J_F}{2},
 \qquad\text{or}\qquad
 R+r
 \quad\longleftrightarrow\quad
 -J_{\bm v^{(2)}}
 =
 \frac{R_H-R_C-J_F}{2}.
 \label{eq:conjectural-4d-3d-Macdonald-charge-map}
\end{equation}
The two choices are exchanged by the Coulomb branch Weyl reflection.

Equation \eqref{eq:conjectural-4d-3d-Macdonald-charge-map} is the central
conjecture of this section.  For $n=2$, it is supported by the explicit
Macdonald-index matching of \cite{Hamachika:2026whv}.  For $n>2$, we do not
have an independent 4d calculation that determines this charge
map.  The evidence for extending it to general $n$ is the uniform
3d charge structure described above.

\subsection{Macdonald refinement}
\label{subsec:Macdonald-refinement}

The conjectural charge map has a direct consequence for the fermionic
representation of the Schur index.  We introduce the two-variable refinement
\begin{equation}
 \begin{aligned}
 {\rm sch}_n^{\mathrm{ref}}(q;y_1,y_2)
 :=
 \sum_{\bm \ell\in\mathbb Z_{\geq0}^{4n-1}}
 \frac{
 q^{\frac12\bm \ell^{\mathsf T}K^{(4n-2)}\bm \ell}
 \bigl(-q^{1/2}\bigr)^{B^{(4n-2)}\cdot\bm \ell}
 y_1^{\bm v^{(1)}\cdot\bm \ell}
 y_2^{\bm v^{(2)}\cdot\bm \ell}
 }{
 \displaystyle\prod_{a=1}^{4n-1}(q;q)_{\ell_a}
 } .
 \end{aligned}
 \label{eq:general-bigraded-Nahm-sum}
\end{equation}
By \eqref{eq:Nahm-grading-topological-charge}, the two fugacities grade
$-J_{\bm v^{(1)}}$ and $-J_{\bm v^{(2)}}$, respectively.  At
$y_1=y_2=1$, this reduces to the unrefined Nahm sum
\eqref{eq:AD-Nahm-sum}.

The exchange $\ell_1\leftrightarrow ell_2$ gives the exact identity
\begin{equation}
 {\rm sch}_n^{\mathrm{ref}}(q;y_1,y_2)
 =
 {\rm sch}_n^{\mathrm{ref}}(q;y_2,y_1),
 \label{eq:general-refined-Weyl-identity}
\end{equation}
and in particular
\begin{equation}
 {\rm sch}_n^{\mathrm{ref}}(q;T,1)
 =
 {\rm sch}_n^{\mathrm{ref}}(q;1,T).
\end{equation}

If the conjectural charge map
\eqref{eq:conjectural-4d-3d-Macdonald-charge-map} holds, the Macdonald index is
obtained by grading the fermionic sum by either of the two Weyl-related
charges.  We therefore predict
\begin{equation}
 \mathcal I_{\mathrm{Mac}}^{(A_2,D_{3n-2})}(q,T)
 =
 {\rm sch}_n^{\mathrm{ref}}(q;T,1)
 =
 {\rm sch}_n^{\mathrm{ref}}(q;1,T).
 \label{eq:general-Macdonald-specialization-conjecture}
\end{equation}
For $n=2$, this is the specialization tested against the known Macdonald
index in \cite{Hamachika:2026whv}.  For $n>2$,
\eqref{eq:general-Macdonald-specialization-conjecture} is a prediction of the
charge-map conjecture.

\subsection{The $n=3$ prediction}
\label{subsec:A10-refinement}

For $n=3$, the associated VOA is $\cA(10)$ and the two charge vectors are
\begin{align}
 \bm v^{(1)}
 &=
 (4,5,1,2,\ldots,9),
 &
 \bm v^{(2)}
 &=
 (5,4,1,2,\ldots,9).
 \label{eq:A10-two-refinement-vectors}
\end{align}
The refined fermionic sum becomes
\begin{equation}
 \begin{aligned}
 {\rm sch}^{\mathrm{ref}}_{\cA(10)}(q;y_1,y_2)
 :=
 \sum_{\bm \ell\in\mathbb Z_{\geq0}^{11}}
 \frac{
 q^{\frac12\bm \ell^{\mathsf T}K^{(10)}\bm \ell}
 \bigl(-q^{1/2}\bigr)^{B^{(10)}\cdot\bm \ell}
 y_1^{\bm v^{(1)}\cdot\bm \ell}
 y_2^{\bm v^{(2)}\cdot\bm \ell}
 }{
 \displaystyle\prod_{a=1}^{11}(q;q)_{\ell_a}
 } .
 \end{aligned}
 \label{eq:A10-bigraded-Nahm-sum}
\end{equation}
Expanding through conformal weight seven gives
\begin{equation}
 \begin{aligned}
 {\rm sch}^{\mathrm{ref}}_{\cA(10)}(q;y_1,y_2)
 ={}&1+y_1y_2q^2+y_1y_2q^3
 +(y_1y_2+y_1^2y_2^2)q^4
 \\
 &+(y_1y_2+y_1^2y_2^2)q^5
 +(y_1y_2+2y_1^2y_2^2+y_1^3y_2^3)q^6
 \\
 &+\bigl(
 y_1y_2+2y_1^2y_2^2+y_1^3y_2^3
 -y_1^4y_2^5-y_1^5y_2^4
 \bigr)q^7
 +O(q^8).
 \end{aligned}
 \label{eq:A10-bigraded-expansion}
\end{equation}

The two negative terms at weight seven are naturally associated with the
doublet generators of $\cA(10)$.  The doublet algebra $\cA(p)$ contains two
generators $a^\pm$ of conformal weight \cite{Feigin:2007sp}
\begin{equation}
 h_{a^\pm}
 =
 \frac{3p-2}{4},
 \label{eq:Ap-doublet-weight}
\end{equation}
so that $h_{a^\pm}=7$ for $p=10$.  These generators are odd
\cite{AdamovicMilas:2013}.  
In the fermionic sum, the contributions with
$\bm \ell=\bm e_1$ and $\bm \ell=\bm e_2$ have bidegrees
\begin{equation}
 \deg(a^+)=(4,5),
 \qquad
 \deg(a^-)=(5,4),
 \label{eq:A10-doublet-bidegrees}
\end{equation}
up to exchanging the two members of the doublet.  This reproduces the two
terms
\begin{equation}
 -y_1^4y_2^5q^7-y_1^5y_2^4q^7
\end{equation}
in \eqref{eq:A10-bigraded-expansion}.

The conjectural charge map then predicts
\begin{equation}
 \mathcal I_{\mathrm{Mac}}^{(A_2,D_7)}(q,T)
 =
 {\rm sch}^{\mathrm{ref}}_{\cA(10)}(q;T,1)
 =
 {\rm sch}^{\mathrm{ref}}_{\cA(10)}(q;1,T),
 \label{eq:A10-Macdonald-conjecture}
\end{equation}
and hence
\begin{equation}
 \begin{aligned}
 \mathcal I_{\mathrm{Mac}}^{(A_2,D_7)}(q,T)
 ={}&1+Tq^2+Tq^3+(T+T^2)q^4+(T+T^2)q^5
 \\
 &+(T+2T^2+T^3)q^6
 +(T+2T^2+T^3-T^4-T^5)q^7
 +O(q^8).
 \end{aligned}
 \label{eq:A10-one-variable-refined-character}
\end{equation}
At $T=1$, this reduces to the unrefined $\cA(10)$ vacuum supercharacter.
Equation \eqref{eq:A10-one-variable-refined-character} is therefore a
quantitative prediction for the 4d Macdonald index following
from the proposed $R$-twisted charge map.

\section{Summary and discussion}
\label{sec:summary-discussion}

We have investigated the family of 4d $(A_2,D_{3n-2})$
Argyres--Douglas theories through the Schur sector SCFT/VOA correspondence
and through candidate 3d theories associated with their R-twisted circle
reductions. On the 4d/2d side, the Schur index agrees to all orders with the
vacuum supercharacter of $\cA(4n-2)$, and the factorization underlying this
identity gives the graded super vector space equivalence described in
Section~\ref{sec:scft-voa}. The conformal gauging construction also suggests
the BRST realization \eqref{eq:BRST-Ap}. For $n>2$, however, the identification
of the full BRST cohomology with $\cA(4n-2)$ remains conjectural.

The fermionic formula for the same character gives a candidate 3d
$\mathcal N=2$ abelian CS matter description and determines the monopole
superpotential \eqref{eq:general-monopole-superpotential} compatible with the
proposed R-twisted reduction. The resulting Coulomb branch is
$\mathbb C^2/\mathbb Z_2$ for every $n$, as in
\eqref{eq:CB-ring-general}, and the superconformal index has the universal
expansion \eqref{eq:n3-SCI-expansion} through order
$\mathfrak q^{3/2}$. In particular, it contains the extra supercurrent terms
expected for enhancement from $\mathcal N=2$ to $\mathcal N=4$.

We also introduced a two variable refinement of the fermionic formula. The
3d charge relation \eqref{eq:physical-topological-charge-identification},
together with the $n=2$ Macdonald index matching, motivates the 4d/3d charge
map \eqref{eq:conjectural-4d-3d-Macdonald-charge-map} and the conjectural
Macdonald index formula
\eqref{eq:general-Macdonald-specialization-conjecture}. For $n=3$, this gives
the explicit prediction \eqref{eq:A10-one-variable-refined-character}, in
which the odd doublet generators of $\cA(10)$ first appear at conformal
weight seven. A direct 4d calculation of the Macdonald index would therefore
provide an independent test of both the proposed formula and the 4d/3d
charge map.

There are three natural directions for further study.  First, the proposed
SCFT/VOA correspondence should be strengthened by computing the BRST
cohomology, including its strong generators and operator products.  Second,
the 3d flow and the $\mathcal N=4$ enhancement should be established beyond
the finite-order index tests, and it would be useful to determine how the
first $n$-dependent operators arise.  Third, the conjectural Macdonald
specialization should be checked directly in four dimensions.  The
$\mathcal N=1$ Lagrangian descriptions of generalized Argyres--Douglas
theories constructed in \cite{Agarwal:2017roi} provide a possible
starting point for such a calculation.  In particular, the family
$(A_{2m},D_{2m(k-1)+k})$ studied there contains the present
$(A_2,D_{3n-2})$ theories for $m=1$ and $k=n$.  It would be interesting to
compute the 4d superconformal index from this Lagrangian description and,
after identifying the fugacities associated with the emergent
$\mathcal N=2$ symmetry, determine whether its Macdonald limit reproduces
\eqref{eq:general-Macdonald-specialization-conjecture}.  Such a calculation
would also give an independent test of
\eqref{eq:conjectural-4d-3d-Macdonald-charge-map}.

\section*{Acknowledgements}
The author is grateful to Takahiro Nishinaka for valuable discussions and for his collaboration at an early stage of this work.

\appendix

\section{Low order magnetic charge sector classification}
\label{app:low-order-sector-classification}

This appendix derives the integrality conditions and the complete list of
magnetic sectors used in Subsection~\ref{subsec:general-n-SCI}.  We keep
$M=4n-4$ and the parametrization
\eqref{eq:general-sector-coordinates}--\eqref{eq:general-sector-components}.
Since the coefficients $c_r$ can be integral or half-integral, it is convenient
to define
\begin{equation}
 C_r:=2c_r,
 \qquad
 C_0:=0,
 \qquad
 C_{M+1}:=s.
 \label{eq:appendix-C-coordinates}
\end{equation}
For an integral magnetic charge, all $C_r$ are integers.  In these variables,
\eqref{eq:general-sector-components} becomes
\begin{align}
 m_1&=a-v,
 &m_2&=a-u,
 \nonumber\\
 2m_{r+2}&=-C_{r-1}+2C_r-C_{r+1},
 &&r=1,\ldots,M,
 \nonumber\\
 2m_N&=2(s-a)-C_M.
 \label{eq:appendix-sector-components-doubled}
\end{align}
The sparse action \eqref{eq:SCI-diagonal-magnetic-basis} also gives
\begin{equation}
 K^{(4n-2)}\bm m
 =
 (a,a,C_1,\ldots,C_M,s).
 \label{eq:appendix-Km-components}
\end{equation}

The conditions $m_{r+2}\in\mathbb Z$ are equivalent to
\begin{equation}
 C_{r-1}\equiv C_{r+1}\pmod 2,
 \qquad r=1,\ldots,M.
 \label{eq:appendix-parity-recurrence}
\end{equation}
Using $C_0=0$, $C_{M+1}=s$, and the fact that $M$ is even, these conditions give
\begin{equation}
 C_{2j}\equiv0\pmod2,
 \qquad
 C_{2j-1}\equiv s\pmod2,
 \qquad
 j=1,\ldots,\frac{M}{2}.
 \label{eq:appendix-parity-solution}
\end{equation}
In particular, $C_M$ is even, so the last component $m_N$ in
\eqref{eq:appendix-sector-components-doubled} is also integral.  Thus
\eqref{eq:appendix-parity-solution} is equivalent to integrality of the full
magnetic charge.  When $s$ is even, all $c_r$ are integers.  When $s$ is odd,
$c_{2j-1}$ are half-integers and $c_{2j}$ are integers.

We next impose the condition $D(\bm m)\leq3$ from
Subsection~\ref{subsec:general-n-SCI}.  For each component, set
\begin{equation}
 d_b
 :=
 \max(m_b,0)-(K^{(4n-2)}\bm m)_b.
\end{equation}
The minimization appearing in \eqref{eq:single-chiral-leading-degree} can be
performed explicitly:
\begin{align}
&\min_{\substack{\rho,\sigma\geq0\\ \rho-\sigma=d_b}}
\left[
 \max(m_b,0)+\rho|m_b|+2\sigma+\sigma|m_b|
 +\sigma(\sigma-1)
\right]
\nonumber\\
&\qquad =
\begin{cases}
 \max(m_b,0)+d_b|m_b|,
 & d_b\geq0,\\[1mm]
 \max(m_b,0)+(-d_b)(|m_b|-d_b+1),
 & d_b<0.
\end{cases}
\label{eq:appendix-local-minimum}
\end{align}
For $d_b\geq0$ the minimum is attained at
$(\rho,\sigma)=(d_b,0)$, while for $d_b<0$ it is attained at
$(\rho,\sigma)=(0,-d_b)$.

At fixed $s$, the contribution associated with $m_{r+2}$ depends only on
$C_{r-1}$, $C_r$, and $C_{r+1}$.  The minimum of $D(\bm m)$ can therefore be
found successively along the chain
\begin{equation}
 C_0=0,\ C_1,\ldots,C_M,\ C_{M+1}=s,
\end{equation}
subject to the parity conditions
\eqref{eq:appendix-parity-solution}.  Substituting
\eqref{eq:appendix-sector-components-doubled} and
\eqref{eq:appendix-Km-components} into
\eqref{eq:appendix-local-minimum} gives, for odd $s$,
\begin{equation}
 D(\bm m)
 \geq
 2n-\frac32+\frac{|s-1|}{2}.
 \label{eq:appendix-odd-s-lower-bound}
\end{equation}
For even $s$, the minimum at fixed $s$ satisfies
\begin{equation}
 D_{\min}(s)
 =
 \frac{s}{2}
 \quad (s\geq0),
 \qquad
 D_{\min}(-2)=3,
 \qquad
 D_{\min}(s)\geq6
 \quad (s\leq-4).
 \label{eq:appendix-even-s-lower-bound}
\end{equation}
Hence, for $n\geq3$, no odd value of $s$ can satisfy $D(\bm m)\leq3$, and the
only allowed values are
\begin{equation}
 s\in\{-2,0,2,4,6\}.
 \label{eq:appendix-allowed-even-s}
\end{equation}

For these five values of $s$, evaluating
\eqref{eq:appendix-local-minimum} under the condition $D(\bm m)\leq3$ gives
$c_r\in\{-1,0,1\}$.  Substitution into
\eqref{eq:appendix-sector-components-doubled} leaves the following solutions:
\begin{align}
 s=-2:&\quad
 (u,v;a;\bm c)=(-1,-1;-1;-\bm1_M),
 \nonumber\\
 s=0:&\quad
 (-1,1;0;\bm0),\ (1,-1;0;\bm0),
 \nonumber\\
 &\quad
 (0,0;0;\bm0),\ (0,0;1;\bm0),\
 (0,0;0;\bm e_j^{(M)}),
 \nonumber\\
 s=2:&\quad
 (-1,3;0;\bm0),\ (3,-1;0;\bm0),
 \nonumber\\
 &\quad
 (u,2-u;0;\bm c),\quad
 u=0,1,2,\quad
 \bm c\in
 \{\bm0,\bm e_1^{(M)},\ldots,\bm e_{M-1}^{(M)}\},
 \nonumber\\
 s=4:&\quad
 (u,4-u;0;\bm0),\quad u=0,\ldots,4,
 \nonumber\\
 s=6:&\quad
 (u,6-u;0;\bm0),\quad u=0,\ldots,6.
 \label{eq:appendix-even-sector-list}
\end{align}
Here $\bm c=(c_1,\ldots,c_M)$,
$\bm e_j^{(M)}$ is the $j$th standard basis vector of $\mathbb Z^M$, and
$\bm1_M=(1,\ldots,1)$.  These are exactly the sectors displayed in
Subsection~\ref{subsec:general-n-SCI}.  Their total number is
\begin{equation}
 1+(M+4)+(3M+2)+5+7=16n+3,
 \qquad n\geq3.
\end{equation}
The classification follows directly from $D(\bm m)\leq3$ and does not require
a cutoff on the components of $\bm m$.

\subsection{Additional sectors at $n=2$}
\label{app:n2-exceptional-sectors}

For $n=2$, one has $M=4$.  The odd-$s$ bound
\eqref{eq:appendix-odd-s-lower-bound} allows one additional value, $s=1$, for
which the minimum is
\begin{equation}
 D_{\min}(1)=\frac52.
\end{equation}
There are eight additional integral magnetic charges with
\begin{equation}
 s=1,
 \qquad
 a=0,
 \qquad
 (u,v)=(0,1)\ \text{or}\ (1,0).
 \label{eq:appendix-n2-exceptional-uv}
\end{equation}
For each choice of $(u,v)$, the allowed values of $\bm c$ and the signs of
their leading contributions are
\begin{equation}
 \begin{array}{c|c}
 (c_1,c_2,c_3,c_4)&\text{sign of the leading term}\\ \hline
 (-\tfrac12,-1,-\tfrac12,0)&-1\\
 (-\tfrac12,0,\tfrac12,0)&+1\\
 (\tfrac12,0,-\tfrac12,0)&+1\\
 (\tfrac12,0,\tfrac12,0)&-1
 \end{array}.
 \label{eq:appendix-n2-exceptional-table}
\end{equation}
All eight sectors have $D(\bm m)=5/2$.  By
\eqref{eq:sector-flavor-weights}, the four sectors with $(u,v)=(0,1)$ have
flavor weight $T^{1/2}x$, while the four sectors with $(u,v)=(1,0)$ have
flavor weight $T^{1/2}x^{-1}$.  For each fixed choice of $(u,v)$, the four
leading coefficients cancel:
\begin{equation}
 -1+1+1-1=0.
 \label{eq:appendix-n2-exceptional-cancellation}
\end{equation}
Therefore the terms of order $\mathfrak q^{5/4}$ cancel separately at
$T^{1/2}x$ and $T^{1/2}x^{-1}$.  The next terms in the one-loop-determinant
expansion occur at least one power of $\mathfrak q$ higher, so these additional
sectors do not contribute through order $\mathfrak q^{3/2}$.  This is why the
universal expression \eqref{eq:n3-SCI-expansion} also holds for $n=2$.

\bibliography{VOA}

\end{document}